\documentclass[11pt]{article}
\usepackage[utf8x]{inputenc}
\usepackage{amsmath, amssymb,bbm}
\usepackage{graphicx,color}
\usepackage{hfstyle}
\newcommand{\mfrac}[2]{\genfrac{}{}{1.0pt}{}{#1}{#2}}

\newcommand{\Pt}[1]{{\mathbf{P}_t^{(#1)}}}
\newcommand{\Ptplus}[1]{{\mathbf{P}_{t+1}^{(#1)}}}
\newcommand{\PP}[1]{{\mathbf{P}^{(#1)}}}

\newcommand{\LL}[1]{\varLambda^{(#1)}}

\begin{document}
\title{Local structure approximation as a predictor of second order phase transitions
in asynchronous cellular automata}
\author{Henryk Fuk\'s$^{1}$, and Nazim Fat\`{e}s$^{2}$}
      \oneaddress{
         $^1$Department of Mathematics\\
         Brock University\\
         St. Catharines, Ontario L2S 3A1, Canada\\   
         \email{\mbox{hfuks@brocku.ca}}\\
         {~}\\
         $^2$ Inria\\
         F-54 600 Villers-l\`{e}s-Nancy, France
         {~} \\
         \email{\mbox{nazim.fates@loria.fr}}
   }

%
\Abstract{
We show that local structure approximation of sufficiently high order can predict the
existence of second order phase transitions belonging to the
directed percolation university class in $\alpha$-asynchronous cellular automata.
}
\maketitle
\section{Introduction}
If the field of deterministic cellular automata is now relatively well known,
their stochastic counterparts remains in great part {\em terra incognita}.
Indeed, even in the simplest case where the systems are binary, one-dimensional,
and where the behaviour of each cell of the automaton depends only on itself and
its two nearest neighbours, little is known on the behaviour of such systems
with random local transitions functions.

As a first step, we focus on this particular set of cellular automata, called
Elementary Cellular Automata (ECA). We consider here {\em $\alpha$-asynchronous
rules}, which are obtained by a random perturbation of the deterministic
updating rule: instead of updating all cells simulatneously, we update each cell
independently with probability~$ \alpha $, the {\em synchrony rate}, and leave
its state unchanged with probability $ 1 -\alpha $.
Two main motivations exist for studying such systems:
\begin{itemize}
\item
In the case where the cellular automaton is used to represent the evolution of a
natural phenomenon, it is interesting to know what is the respective role of the
local rule and the updating procedure in the outcome of a
simulation~\cite{IngBuv84,Sch99}. The study of a continuous variation of $
\alpha $ for a quasi-deterministic setting ($\alpha \rightarrow 1$) to a
quasi-sequential one ($\alpha \rightarrow 0$) may allow us to detect the
non-robustness of some systems (see e.g. the study of Grilo and Correia on the
iterated prisoner's dilemma~\cite{GriCor11}). 
\item $\alpha$-asynchronous cellular automata can also be seen as the result of
a stochastic ``mixing'' of two deterministic rules (the original rule and the
identity rule). 
However, in contrast to other perturbations where the outcome of the mix can be
partially predicted (e.g., mixing a rule with uniform noise or with a ``null''
rule),  the effect of switching from the deterministic setting ($\alpha=1$) to a
probabilistic one ($\alpha<1$) is to date unknown, even if $ \alpha $ is
infinitely close to $ 1 $. The origin of this fact can be intuitively perceived
by considering that, from the point of view of each single cell, all happens as
if it were updated with an independent clock where the time between two updates
follows a geometric law of parameter $ \alpha $. It is therefore difficult, if
not impossible, to predict in all generality whether the introduction of
asynchrony will stabilize or destabilize the system~\cite{BerDet94,FMST06}. 
\end{itemize}

The systematic exploration of the properties of $ \alpha$-asynchronous
Elementary Cellular Automata by numerical simulations revealed that different
``responses'' to this perturbation were observed: some rules, as the majority
rule (ECA 232), show only little change while other rules (e.g., ECA~2) show a
drastic modification of their behaviour as soon as a little amount of
asynchronism is introduced~\cite{FatMor05}.
However, the most surprising phenomenon was the identification of rules  
which exhibited a qualitative change of behaviour for a continuous variation of
the synchrony rate: there exists a critical value of $ \alpha_c $ which separates
a an {\em active phase} in which the system fluctuates around en equilibrium and
an {\em absorbing phase} where the system is rapidly attracted towards a fixed
point where all cells are in the same state.

Using the techniques from statistical physics, this abrupt change of behaviour
was then identified as a second order phase 
transitions which belong to the directed percolation (DP) universality
class~\cite{FatJCA09}. 
This identification was conducted by taking as an order parameter the density,
that is the average number of cells in state 1, and, up to symmetries, nine
rules were found to exhibit such DP behaviour. Their Wolfram numbers is:
6, 18, 26, 38, 50, 58, 106, 134, and 146.

The aim of this paper is to study to which extent this second order phase
transition can be predicted with analytical techniques. We are in particular
interested in answering two questions: (a) Can we explain the existence of the
two active and absorbing phases? (b) Can we propose an approximation of the
value of the critical synchrony rate $ \alpha_c $ that separates the two phases?

Our approach is based on so-called \emph{local structure theory}, proposed
in 1987 by  H. A. Gutowitz \emph{et al.} \cite{gutowitz87a,gutowitz87b} as a
generalization
of the mean-field theory for cellular automata. Unlike mean-field theory,  local
structure theory takes (partially) into account 
correlations between sites. The basic idea of this theory is to consider
probabilities of blocks (words) of length $k$ and to construct a map on these
block probabilities, which, when iterated,
approximates probabilities of occurrence of the same  blocks in the actual orbit
of a given  cellular automaton.
The construction is based on  the idea of  ``Bayesian extension'', introduced
earlier by other  authors
in the context of lattice gases
\cite{Brascamp71,Fannes84}, and also known as a ``finite-block measure'' or as
``Markov process with memory''.
Although Gutowitz \emph{et al.} originally considered deterministic CA rules,
extension to probabilistic rules
is straightforward, and has been described in detail in \cite{paper50}.
In the case of nearest-neighbour binary rule, the aforementioned map is
$2^k$-dimensional, where $k$ is called 
the \emph{level} of local structure approximation. However, using the method
proposed in \cite{paper50},
it can be reduced to equivalent, but somewhat simpler $2^{k-1}$ dimensional map,
and we will fully exploit this simplification
here.

It has been observed that for many CA rules,  as the level $k$ increases, the
accuracy of the approximation
increases as well. More importantly, in some PCA,  their local structure map
``inherits'' important features of the PCA. For example, Mendon\c{c}a  and de Oliveira \cite{Mendoca2011}
studied a PCA which can be understood as a ``probabilistic mixture'' of 
elementary CA rules 182 and 200, such that at a given site, rule 182
is applied with a probability $p$, and rule 202 with probability $1-p$. They found that
as $p$ varies, the density of ones in the steady state undergoes a phase transition which,
according to numerical evidence, belongs to DP (directed percolation) universality class.
They also found that the mean-field approximation of this rule predicts existence of
this phase transition, meaning that the mean field map exhibits a bifurcation
with exchange of stability between two fixed points. For higher level local structure approximation 
(level 2, 3 and 4),
the authors obtained density curves approximating the actual density curve with increasing
accuracy as the level increased.

In the rest of this paper, we will demonstrate that for $ \alpha$-asynchronous
Elementary Cellular Automata exhibiting phase transitions
belonging to DP universality class this is also the case: the local structure approximation
not only predicts existence of the phase transition, but does that increasingly well
as the level of the approximation increases.

\section{Probabilistic cellular automata}
We will assume that the dynamics takes place on a one-dimensional lattice. Let
$s_i(t)$ denotes the state of the lattice site $i$ at time $t$, where
$i \in \mathbb{Z}$, $t \in \mathbb{N}$.  We will
further assume that $s_i(t) \in \{ 0, 1 \}$ and we will say that the
site $i$ is occupied (empty) at time $t$ if $s_i(t)=1$ ($s_i(t)=0$).
\emph{Deterministic elementary cellular automaton} is the dynamical system 
governed by the local function $f: \{0,1\}^3 \to \{0,1\}$ such that
$$s_i(t+1)=f\big(s_{i-1}(t), s_{i}(t), s_{i+1}(t)\big).$$ Function $f$ is be
called
a \emph{rule} of CA.

In a probabilistic cellular automaton (PCA), lattice sites
simultaneously change states form $0$ to $1$ or from $1$ to $0$
with probabilities depending on states of local neighbours. A
common method for defining PCA is to specify a set of local
transition probabilities. For example, in order to define a
nearest-neighbour PCA one has to specify the probability
$w(s_i(t+1))| s_{i-1}(t),s_i(t),s_{i+1}(t))$ that the site
$s_i(t)$ with nearest neighbors $s_{i-1}(t),s_{i+1}(t)$ changes
its state to $s_i(t+1)$ in a single time step.

A more formal definition of nearest-neighbour PCA can be constructed as follows.
Consider  a set of independent Boolean random variables
$X_{i,\mathbf{b}}$, where $i \in \mathbb{Z}$ and $\mathbf{b} \in
\{0,1\}^3$. Probability that the random variable
$X_{i,\mathbf{b}}$ takes the value $a \in\{0,1\}$ will be
assumed to be independent of $i$ and
 denoted
by $w(a|\mathbf{b})$, 
\begin{equation}\label{defw}
Pr(X_{i,\mathbf{b}}=a)=w(a|\mathbf{b}).
\end{equation}
Obviously, $w(1|\mathbf{b})+w(0|\mathbf{b})=1$ for all $\mathbf{b}
\in \{0,1\}^n$. The update rule for PCA is then defined by
\begin{equation}\label{defprobca}
  s_i(t+1)=X_{i,\{s_{i+l}(t),s_{i+l+1}(t),\ldots,s_{i+r}(t)\}}.
\end{equation}
Note that new random variables $X$ are used at each time step $t$,
that is, random variables $X$ used at time step $t$ are
independent of those used at previous time steps. 

With the above definition, it is clear that in order to fully define a nearest-neighbour PCA rule,
it is enough to specify eight transition probabilities $w(1|x_1x_2x_3)$ for all $x_1,x_2,x_3 \in \{0,1\}$. 
Remaining eight probabilities, $w(0|x_1x_2x_3)$, can be obtained by $w(0|x_1x_2x_3)=1-w(1|x_1x_2x_3)$.

We will now define \emph{$\alpha$-asynchronous elementary cellular automata}. Let $\alpha \in [0,1]$ and let $f$ be a local function
of some deterministic CA with Wolfram number $W(f)$.  Corresponding $\alpha$-asynchronous elementary cellular automaton
with rule number $W(f)$ is a probabilistic CA for which transition probabilities are
\begin{equation}
 w(1|x_1x_2x_3)=\alpha f(x_1,x_2,x_3)+(1-\alpha)x_2.
\end{equation}
Note that when $\alpha=1$, the above becomes just the deterministic rule with local function $f$, and
when $\alpha=0$, it becomes the identity rule.

\section{Local structure approximation}
In what follows we assume that the probabilistic CA rule is binary and that the neighbourhood
size is three (central site and two nearest neighbours). It is not difficult, however, to generalize these
results to CA with higher number of states and larger neighbourhood.

We denote by $ P_t(\mathbf{b})$
the probability 
of occurrences of blocks $\mathbf{b}=b_1b_2 \ldots b_n$ after $t$ iterations of the PCA rule, where
$\mathbf{b} \in \{0,1\}^*$, that is, $\mathbf{b}$ is a word over the binary alphabet.
More precisely,
\begin{equation}
P_t(\mathbf{b})=Pr\Big(s_i(t)=b_1, s_{i+1}(t)=b_2, \ldots s_{i+n-1}(t)=b_n \Big),
\end{equation}
where we assume that  $P_t(\mathbf{b})$ is independent of $i$, since we will be only interested
in shift-invariant states.

One should add at this point that the {\em block probabilities} $P_t(\mathbf{b})$, as we will call them,
are more formally measures  of  cyllinder sets  with respect to a
shift-invariant measure on $\{0,1\}^\mathbb{Z}$. Review of all details of the construction of this
measure can be found in \cite{paper50}, thus we will not discuss these details here.
We will only remark that the knowledge of \emph{all} block probabilities is equivalent
to the knowledge of probability measure on $\{0,1\}^\mathbb{Z}$, by the virtue of Hahn-Kolmogorov
extension theorem~\cite{paper50}. Therefore, the sequence of sets of block probabilities 
$$\{P_t(\mathbf{b}): \,\, \mathbf{b}\in \{0,1\}^{\star} \}$$
with $t=0, 1, 2 \ldots$ can be viewed as a sequence of probability measures 
on $\{0,1\}^\mathbb{Z}$. Moreover, if $P_t(\mathbf{b})=P_{t+1}(\mathbf{b})$ for all $\mathbf{b}$,
then these block probabilities define \emph{invariant measure}. It has been observed that in many
PCA rules, as $t \to \infty$, block probabilities tend to some stationary or ``equilibrium'' value.
These stationary block probability values will be denoted by $P(\mathbf{b})$, that is,
without index $t$. Obviously, they extend to invariant measures. 

In some cases, for blocks $\mathbf{b}$ of short lenght,
one can calculate $P_t(\mathbf{b})$ directly, as, for example, has been done for $\alpha$-asynchronous
ECA rules 76, 140 and 200 in \cite{paper44}. For $\alpha$-asynchronous rules exhibiting phase transitions
such direct calculations are not possible, thus we will use approximate method known as
local structure theory.

Block probabilities form an infinite hierarchy
\begin{center} 
$P_t(0), P_t(1)$\\
$P_t(00),P_t(01),P_t(10),P_t(11)$\\
$P_t(000),P_t(001),P_t(010),P_t(011),P_t(100),P_t(101),P_t(110),P_t(111)$\\
$\cdots$
\end{center}
that we can arrange by 
defining $\mathbf{P}_t^{(k)}$ as a column vector that holds all the $k$-block probabilities sorted in lexical order, that is,\\
\begin{align*}
 \mathbf{P}_t^{(1)}&=[P_t(0), P_t(1)]^T,\\
\mathbf{P}_t^{(2)}&=[P_t(00),P_t(01),P_t(10),P_t(11)]^T,\\
\mathbf{P}_t^{(3)}&=[P_t(000),P_t(001),P_t(010),P_t(011),P_t(100),P_t(101),P_t(110),P_t(111)]^T,\\
&\cdots .
\end{align*}
Components of $\mathbf{P}_t^{(k)}$ are not independent: 
they obey relationships known as
{\em consistency conditions}, which have the form
\begin{equation}
 P_t(\mathbf{b})=P_t(\mathbf{b}0)+P_t(\mathbf{b}1)=P_t(0\mathbf{b})+P_t(1\mathbf{b})
\end{equation}
for any block $\mathbf{b}$. 
These consistency conditions imply that generally only half of components of  $\mathbf{P}_t^{(k)}$ are independent~\cite{paper50}.

Let us now suppose that a PCA is given, and we know its transition probabilities $w$. Let $\mathbf{a}=a_1a_2\ldots a_k$ and $\mathbf{b}=b_1b_2\ldots b_{k+2}$ be two words of size $ k $ and $ k+2 $, respectively, 
then the probability that blocks  $\mathbf{a}$ results from an application of the local
 rule to block $\mathbf{b}$ is given by
\begin{eqnarray}\label{defw2}
w(\mathbf{a}|\mathbf{b})=\prod_{i=1}^k w(a_i|b_ib_{i+1}b_{i+2}),
\end{eqnarray}
where we took advantage of the fact that  cells are independent.
We can thus write:
\begin{equation}\label{blockevol}
P_t(\mathbf{a}) = \sum_{\mathbf{b}\in \{0,1\}^{k+2}} w(\mathbf{a} | \mathbf{b}) P_t(\mathbf{b}),
\end{equation}
or, equivalently in matrix notation
\begin{eqnarray*}
\Ptplus{k} = W^{(k)} \Pt{k+2},
\end{eqnarray*}
where $W^{(k)}$ is a binary matrix with $2^{k}$ rows and $2^{k+2}$ columns with entries given by eq.~(\ref{defw2}).

If an invariant measure exists, then it is given by the set of block probability vectors $\PP{k}$,
$k \in \mathbb{N}$, satisfying
\begin{eqnarray*}
\PP{k}=W^{(k)} \PP{k+2}.
\end{eqnarray*}
Thus, if we want to know what are the block probabilities for blocks
of length $k$, we need to solve the above equation. The problem is that
to know $\PP{k}$, one needs to know $\PP{k+2}$.

One possible solution is to approximate $\mathbf{P}^{(k+2)}$ by expressing it in
terms of $\mathbf{P}^{(k)}$.
Such approximation is known as
Bayesian extension \cite{gutowitz87a}, and is given by
\begin{equation}\label{bayes}
  P(b_1b_2\ldots b_{k+2}) \approx \begin{cases}
                         P(b_1)P(b_2)P(b_3)    & \text{if $k=1$},\\[0.4em]
                      \displaystyle     \frac{P(b_1 \ldots b_{k}) P(b_2 \ldots b_{k+1}) P(b_3 \ldots b_{k+2}) }{P(b_2 \ldots b_{k}) P(b_3\ldots b_{k+1}) }
  & \text{if $k>1$}.
                            \end{cases}
\end{equation}
where we assume that the denominator is positive. If the denominator is zero, then
we take  $ P(b_1b_2\ldots b_{k+2}) =0$.
In order to avoid writing separate cases for denominator equal to zero, we define  ``thick bar'' fraction as
\begin{equation} \label{convention1}
 \mfrac{a}{b}:=\begin{cases}
               {\displaystyle \frac{a}{b}} & \mathrm{if\,\,} b \neq 0\\[1em]
                   0 & \mathrm{if\,\,} b = 0.      
                     \end{cases}
\end{equation}
Moreover, in order to avoid writing the $k=1$ case separately, we adopt notational convention that 
\begin{equation} \label{convention2}
 P(a_m \ldots a_n) =1 \mathrm{\,\, whenever \,\,} n<m,
\end{equation}
and then we can write eq. (\ref{bayes}) simply as
\begin{equation} \label{bayes2}
  P(b_1b_2\ldots b_{k+2}) \approx  \mfrac{P(b_1 \ldots b_{k}) P(b_2 \ldots b_{k+1}) P(b_3 \ldots b_{k+2}) }{P(b_2 \ldots b_{k}) P(b_3\ldots b_{k+1}) },
\end{equation}
which remains valid even for $k=1$.

The numerator of the fraction on the right hand side of eq.~(\ref{bayes2}) contains only blocks
of length $k$, and the denominator only blocks of length $k-1$. We can, however,
express $k-1$ blocks by $k$ blocks using consistency conditions,
\begin{align}
 P(b_2 \ldots b_{k})&=P(b_2 \ldots b_{k}0)+P(b_2 \ldots b_{k}1),  \\
 P(b_3\ldots b_{k+1})&=P(b_3\ldots b_{k+1}0)+P(b_3\ldots b_{k+1}1).
\end{align}
With the approximation given in eq. (\ref{bayes2}), for $k>1$, eq. (\ref{blockevol}) becomes
 \begin{multline}\label{lsteq}
P_{t+1}(a_1 \ldots a_{k})=\\ \sum_{b\in \{0,1\}^{k+2}}
\prod_{i=1}^k w(a_i|b_ib_{i+1}b_{i+2}) \mfrac{P_t(b_1 \ldots b_{k}) P_t(b_2 \ldots b_{k+1}) 
P_t(b_3 \ldots b_{k+2}) }
{
\big( P_t(b_2 \ldots b_{k}0)+P_t(b_2 \ldots b_{k}1)  \big)
\big( P_t(b_3\ldots b_{k+1}0)+P_t(b_3\ldots b_{k+1}1) \big) 
},
\end{multline}
and it should be understood as a system of $2^k$ equations, so that we have a separate equation  for each 
$(a_1 \ldots a_{k}) \in \{0,1\}^k$. In vector form we will write
\begin{equation}
\mathbf{P}^{(k)}_{t+1}= \varLambda^{(k)} \left( \mathbf{P}^{(k)}_t\right),
\end{equation}
where $\varLambda^{(k)}$, defined by eq. (\ref{lsteq}), will be called \emph{local structure map}
of level $k$. As we will see, for $\alpha$-asynchronous cellular automata, behaviour of
fixed points of the local structure maps as a function of the synchrony rate
$\alpha$ can be used as a predictor of existence of phase transitions.

We already mentioned that due to consistency conditions,
not all components of the block probability vector  $\mathbf{P}_t^{(k)}$
are independent. For example, for $k=3$, it is sufficient to consider
only first four components of $\mathbf{P}_t^{(3)}$, that is,
$P_t(000),P_t(001), P_t(010), P_t(011)$. The remaining four can be expressed as
\begin{align*}
P_t(100) &= P_t(001), \\  
P_t(101) &= -P_t(001)+P_t(010)+P_t(011), \\ 
 P_t(110) &= P_t(011), \\  
P_t(111) &= 1-P_t(000)-P_t(001)-2 P_t(010)-3 P_t(011).
\end{align*}
These substitutions can be used to reduce $\LL{3}$, the eight-dimensional local structure map of level 3, to a four-dimensional map. 
The same can be done for arbitrary $k$.
Although $\LL{3}$ has $2^k$ components, not all of them are independent, and  with the help of consistency conditions it can be
reduced to $2^{k-1}$ components~\cite{paper50}.

For $k=1$, eq.~(\ref{blockevol}) takes a simpler form,
\begin{equation}\label{mfeq}
P_{t+1}(a)= \sum_{\mathbf{b}\in \{0,1\}^{3}}
w(a|b_1b_2b_3) P(b_1)P(b_2) P(b_3), 
\end{equation}
where $a \in \{0,1\}$. This defines the two-dimensional map $\varLambda^{(1)}$,
\begin{equation}
\mathbf{P}^{(1)}_{t+1}= \varLambda^{(1)} \left( \mathbf{P}^{(1)}_t\right),
\end{equation}
which we will call the \emph{mean-field map}. 
Although this is an $\mathbb{R}^2 \rightarrow \mathbb{R}^2$ map, again consistency conditions reduce it to
one-dimensional map. Mean-filed map can be considered a special case of local structure map corresponding to $k=1$
(that is, the first-level of local structure approximation).

For $\alpha$-asynchronous rules, 
\begin{align}
w(1|b_1b_2b_3)&=\alpha f(b_1,b_2,b_3) +(1-\alpha) b_2,\\
w(0|b_1b_2b_3)&=1-w(1|b_1b_2b_3), 
\end{align}
where  $b_1,b_2,b_3 \in \{0,1\}$. Equation (\ref{mfeq}) for $a=1$ thus takes the form
\begin{align}
 P_{t+1}(1)= \sum_{b_1,b_2,b_3\in \{0,1\}^{3}} 
\big(
\alpha f(b_1,b_2,b_3) +(1-\alpha) b_2 \big)P_t(b_1)P_t(b_2) P_t(b_3).
\end{align}
Note that the second term will be non-zero only when $b_2=1$, and that summations of the second term over $b_1$ and $b_3$
yield factor $1$ due to consistency conditions $P_t(0)+P_t(1)=1$. We obtain, therefore,
\begin{align}
 P_{t+1}(1)= \alpha \sum_{b_1,b_2,b_3\in \{0,1\}^{3}} f(b_1,b_2,b_3)P_t(b_1)P_t(b_2) P_t(b_3) +(1-\alpha) P_t(1).
\end{align}
In the above, the sum $\sum f(b_1,b_2,b_3) P_t(b_1)P_t(b_2) P_t(b_3)$ will be a function of $P_t(1)$ 
and $P_t(0)$, but by substitution $P_t(1)=1-P_t(0)$ we can convert it to a function of $P_t(1)$ only,
to be denoted $G(P_t(1))$, yielding
\begin{align}
 P_{t+1}(1)= \alpha G(P_t(1)) +(1-\alpha) P_t(1),
\end{align}
or equivalently 
\begin{align}
 P_{t+1}(1)= P_t(1)+\alpha \big(G(P_t(1)) - P_t(1)\big).
\end{align}
The above mean-field map of $\alpha$-asynchronous rule has fixed points given by
\begin{equation}
 G(P_t(1)) - P_t(1)=0,
\end{equation}
and it is clear that they do not depend on $\alpha$. This means, in particular, that the 
mean-field map of $\alpha$-asynchronous rule has the same fixed
point(s) as the mean-field map of the corresponding synchronous rule (for which $\alpha=1$).

\section{Transcritical bifurcation}
\begin{figure}[b!]\label{logistic}
 \begin{center}
     \includegraphics[scale=0.3]{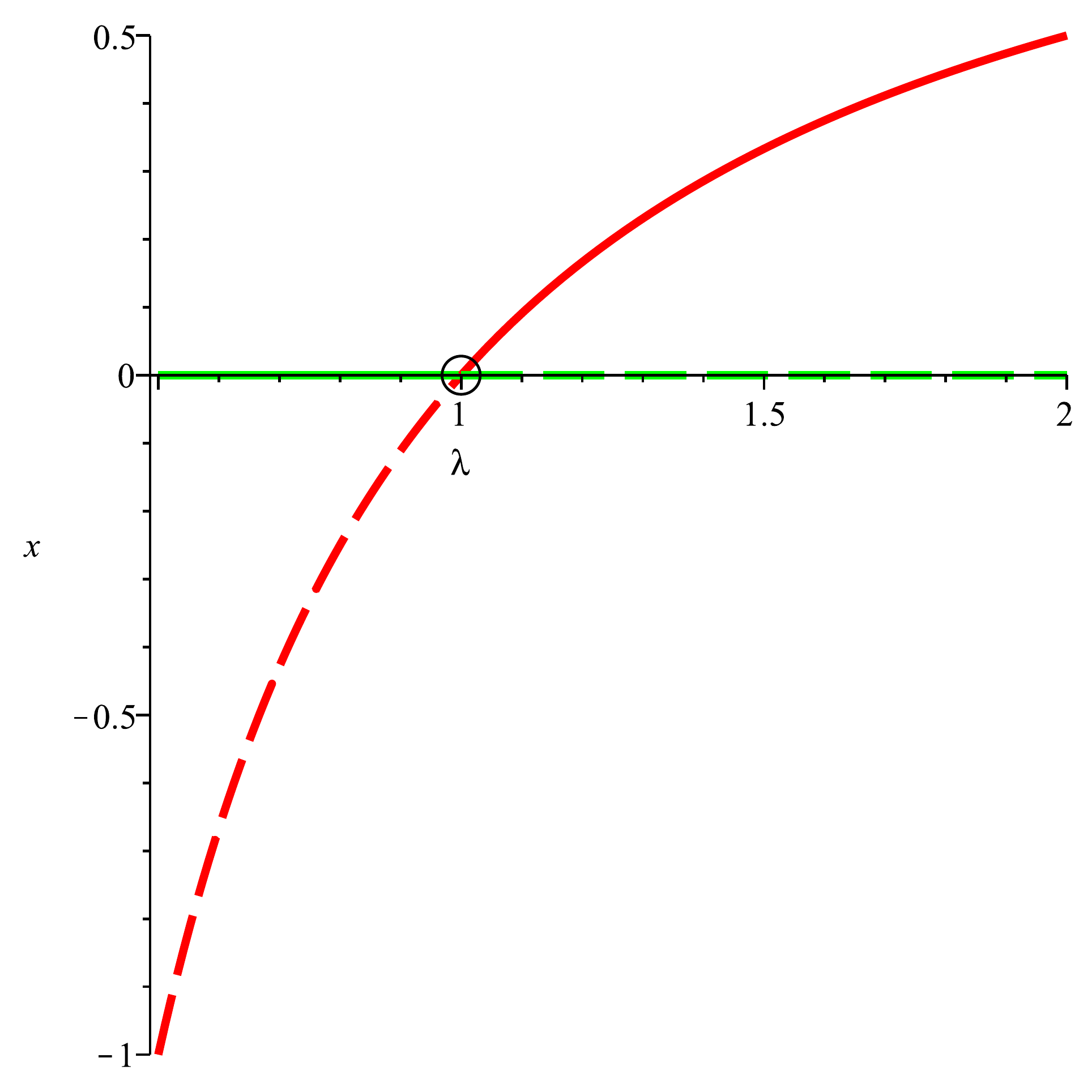}
   \end{center}
\caption{Bifurcation diagram for the logistic map. Solid line represents stable fixed point,
 dashed line unstable fixed point. 
Transcritical bifurcation point at $\lambda=1$ is circled.}
\end{figure}

Before we continue, a brief interjection regarding transcritical bifurcations
would be in place.
As an example, let us consider the well-known logistic map
\begin{equation}
 x \rightarrow \lambda x (1-x),
\end{equation}
where $x\in [0,1]$. The parameter $\lambda$  is normally assumed to be in the interval $[0,4]$,
but we will be interested in only a smaller interval, say $\lambda \in [1/2,2]$.
The logistic map has two fixed points, $x^-=0$ and $x^+=(\lambda-1)/\lambda$. 
It is easy to show that for $\lambda<1$, $x^-$ is stable and $x^+$ is unstable, whereas for $\lambda>1$ their roles reverse, 
that is,
$x=0$ is unstable and $x=(\lambda-1)/\lambda$ is stable. At $\lambda=1$ we thus observe an exchange of stability
between fixed points, and such phenomenon is known as a \emph{transcritical bifurcation}.
Figure~\ref{logistic} shows the bifurcation diagram of the logistic map in the interval
 $[1/2,2]$, that is, 
 graphs of both fixed points as a function of $\lambda$.
Exchange of stability between two fixed points can also happen in higher-dimensional maps,
including local structure maps,
as we will shortly see below.
\section{Local structure approximation for  $\alpha$-ECA 6}
The first example we are going to consider is $\alpha$-asynchronous elementary cellular automaton with rule code 6, defined by
\begin{align}
w(1| 000) &= 0, \,\,\, w(1| 001) = \alpha, \,\,\, w(1| 010) = 1, \,\,\, w(1| 011) = 1-\alpha, \\
w(1| 100) &= 0, \,\,\, w(1| 101) = 0, \,\,\, w(1| 110) = 1-\alpha, \,\,\, w(1| 111) = 1-\alpha.
\end{align}
For $k=1$, that is, for the mean-field approximation, we obtain, from eq. (\ref{mfeq}),
\begin{align}
P_{t+1}(0)&= P_t(0)^3+(2-\alpha) P_t(1) P_t(0)^2+(1+2 \alpha) P_t(1)^2 P_t(0)+\alpha P_t(1)^3,\\
P_{t+1}(1)&= (1+\alpha) P_t(1) P_t(0)^2+2(1- \alpha) P_t(1)^2 P_t(0)+(1-\alpha) P_t(1)^3.
\end{align}
Consistency condition $P_t(0)+P_t(1)=1$ allows to eliminate one of the variables, thus we obtain 
the one-dimensional map
\begin{equation}
P_{t+1}(1) =(1+\alpha) P_t(1)-4 \alpha P_t(1)^2+2 \alpha P_t(1)^3.
\end{equation}
Its fixed points are $0, 1+\frac{\sqrt{2}}{2}, 1-\frac{\sqrt{2}}{2}$, and among them, only
$0$ is in the interval $[0,1]$, thus it is the only admissible fixed point. Since this fixed point
is independent of $\alpha$, the mean field obviously does not exhibit any bifurcation, and we have to consider 
higher order of the local
structure approximation.

For $k=2$, we have $2^2=4$ local structure equations given by eq. (\ref{lsteq}).
Assuming that denominators are not zero, and replacing the ``thick bar'' by 
a regular one,  using variables
$x_t=P_t(00)$, $y_t=P_t(10)$, $z_t=P_t(10)$, and $v_t=P_t(11)$,  eqs. (\ref{lsteq})  
after simplification become
\begin{multline}
x_{t+1}=\frac{1}{(z_t+v_t) (x_t+y_t)^{2}}
\{v_t ( x_t+y_t )  ( v_tx_t+y_tv_t+y_t^2 ) {\alpha}^2\\+
 ( -x_ty_t{z_t}^2+2 y_tv_tz_tx_t-z_tx_t^2y_t+z_tx_t^2v_t+x_ty_t^2v_t+2 y_t^2v_tz_t
 ) \alpha \\
+x_t ( z_t+x_t )  ( x_t+y_t )  ( z_t+v_t
 )\},
\end{multline}
\begin{multline}
y_{t+1}= \frac{1}{(z_t+v_t) (x_t+y_t)^{2}}
\{-v_t ( x_t+y_t )  ( v_tx_t+y_tv_t+y_t^2 ) {\alpha}^2 \\ +
 ( x_ty_t{z_t}^2-y_t^2v_tz_t-z_tx_ty_t^2+2 x_ty_tv_t^2+y_t^3v_t+v_t^2x_t
^2+y_t^2v_t^2 ) \alpha
\\+y_t ( z_t+x_t )  ( x_t+y_t
 )  ( z_t+v_t ) \},
\end{multline}
\begin{equation}
z_{t+1}=  \frac{ -v_t ( v_tx_t+y_tv_t+y_t^2 ) {\alpha}^2+v_t ( x_t+y_t ) 
 ( -z_t+y_t+v_t ) \alpha+z_t ( x_t+y_t )  ( y_t+v_t
 ) }{(z_t+v_t) (x_t+y_t)}, \mbox{\hspace{1.4cm}}
\end{equation}
\begin{multline}
v_{t+1}=  \frac{ v_t ( v_tx_t+y_tv_t+y_t^2 ) {\alpha}^2+ ( x_ty_tz_t-2 y_t^2v_t-2
 v_t^2x_t-2 y_tv_t^2-x_ty_tv_t ) \alpha+v_t ( x_t+y_t ) 
 ( y_t+v_t ) }{(z_t+v_t) (x_t+y_t)}.
\end{multline}
Because of consistency conditions, only first two variables are independent, and the remaining ones
can be expressed by the first two as
\begin{align}
z_t&=P_t(10)=P_t(01)=y_t, \\
v_t&=P_t(11)=1-P_t(00)-P_t(01)-P_t(10)=1-x_t-2y_t.
\end{align}
This allows to reduce the original system to only two equations, with variables $x$ and $y$. 
We can then find  fixed point of this system by dropping indices $t$ and $t+1$, and solving it
for $x$ and $y$. 
This can be done with the help of a symbolic algebra software, yielding 
\begin{align}
x= P(00) &= \frac{\alpha+1-(1-\alpha)\sqrt{4\alpha+1}}{2\alpha(2-\alpha)}, \\
y= P(01) &= \frac{\alpha-1+(1-\alpha)\sqrt{4\alpha+1}}{2\alpha(2-\alpha)}, \\
z= P(10) &= \frac{\alpha-1+(1-\alpha)\sqrt{4\alpha+1}}{2\alpha(2-\alpha)}, \\
v= P(11) &= \frac{\alpha-2\alpha^2+1-(1-\alpha)\sqrt{4\alpha+1}}{2\alpha (2-\alpha)}.
\end{align}
It is easy to check that this further yields
\begin{equation}
 P(1)=1-P(0)=1-(P(00+P(01))=1-x-y=\frac{1-\alpha}{2-\alpha},
\end{equation}
which remains positive for $\alpha\in [0,1)$. One can, moreover, show that the above fixed point is 
always stable. 
Because of the ``thick bar'' convention, one also demonstrate that equations (\ref{lsteq})
have in the case of $k=2$ also another fixed point, $x=1$, $y=z=v=0$. One can show that it is always unstable.
This means that the local structure map of level 2 does not undergo any
bifurcation, and that the local structure theory of level 2 does not
predict any abrupt change in the density of ones as $\alpha$ changes. We need, therefore, to 
consider level 3 approximation.

\begin{figure}
 \begin{center}
Rule 6 \hspace{8cm} Rule 18\\
\includegraphics[scale=0.35]{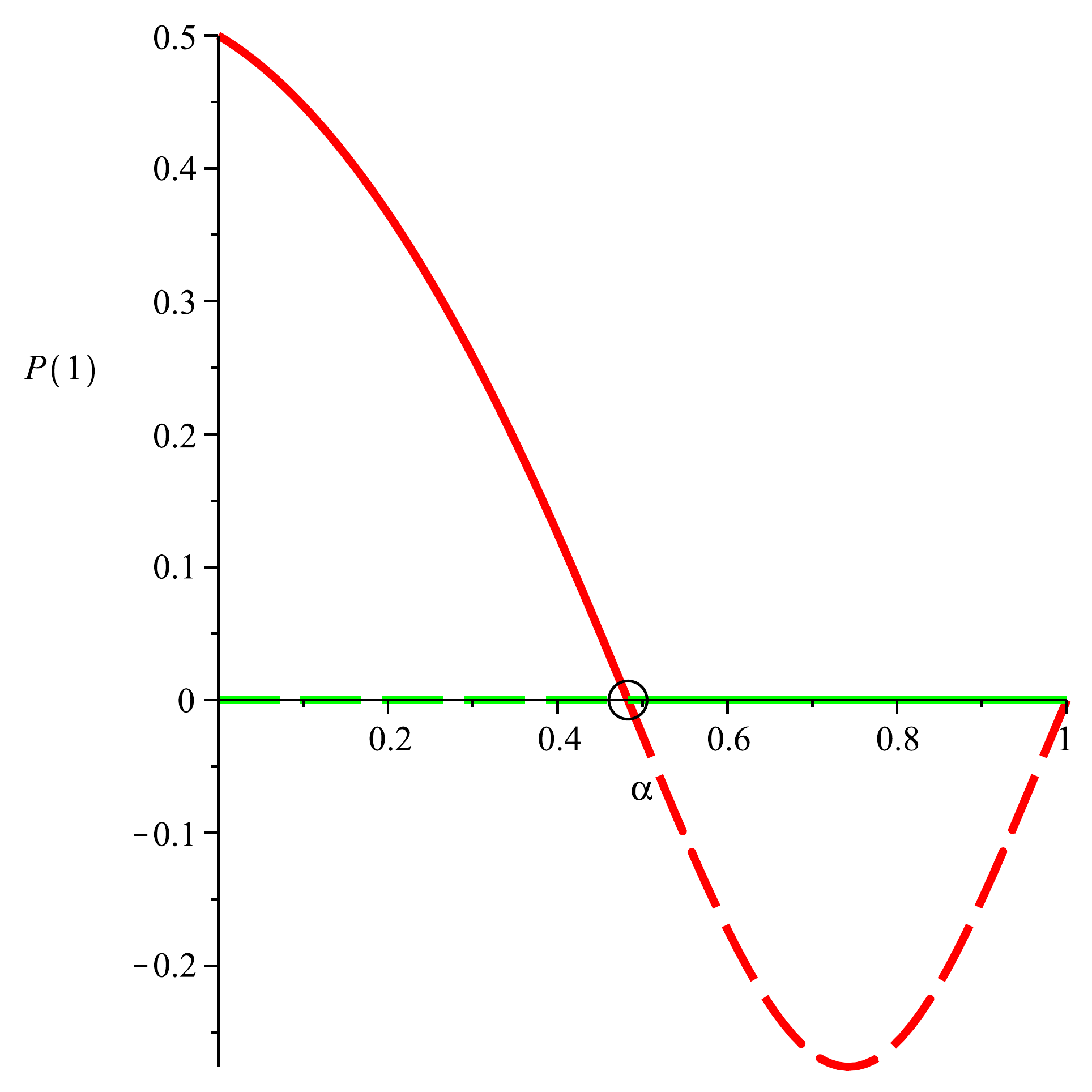} \includegraphics[scale=0.35]{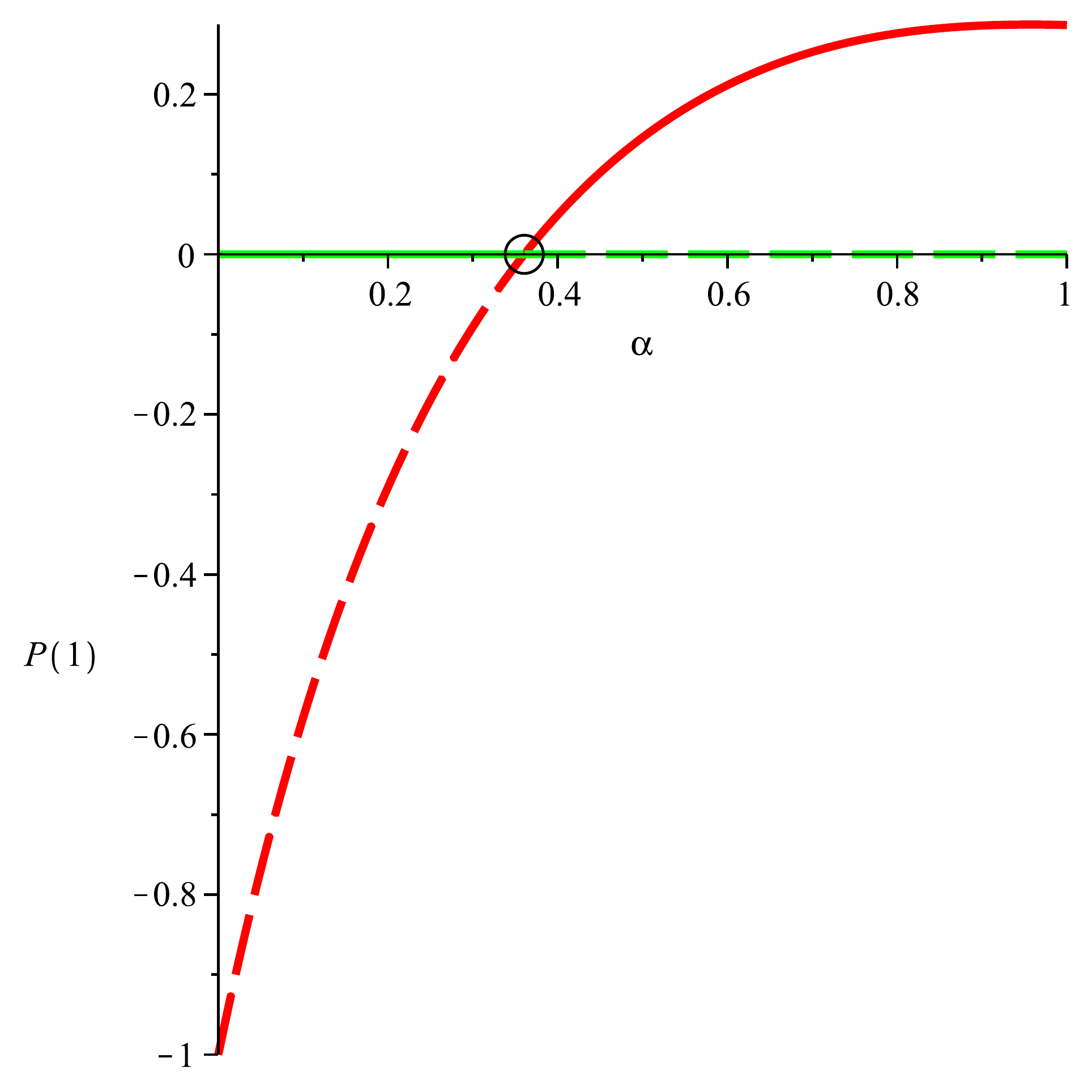}\\
 \includegraphics[scale=0.35]{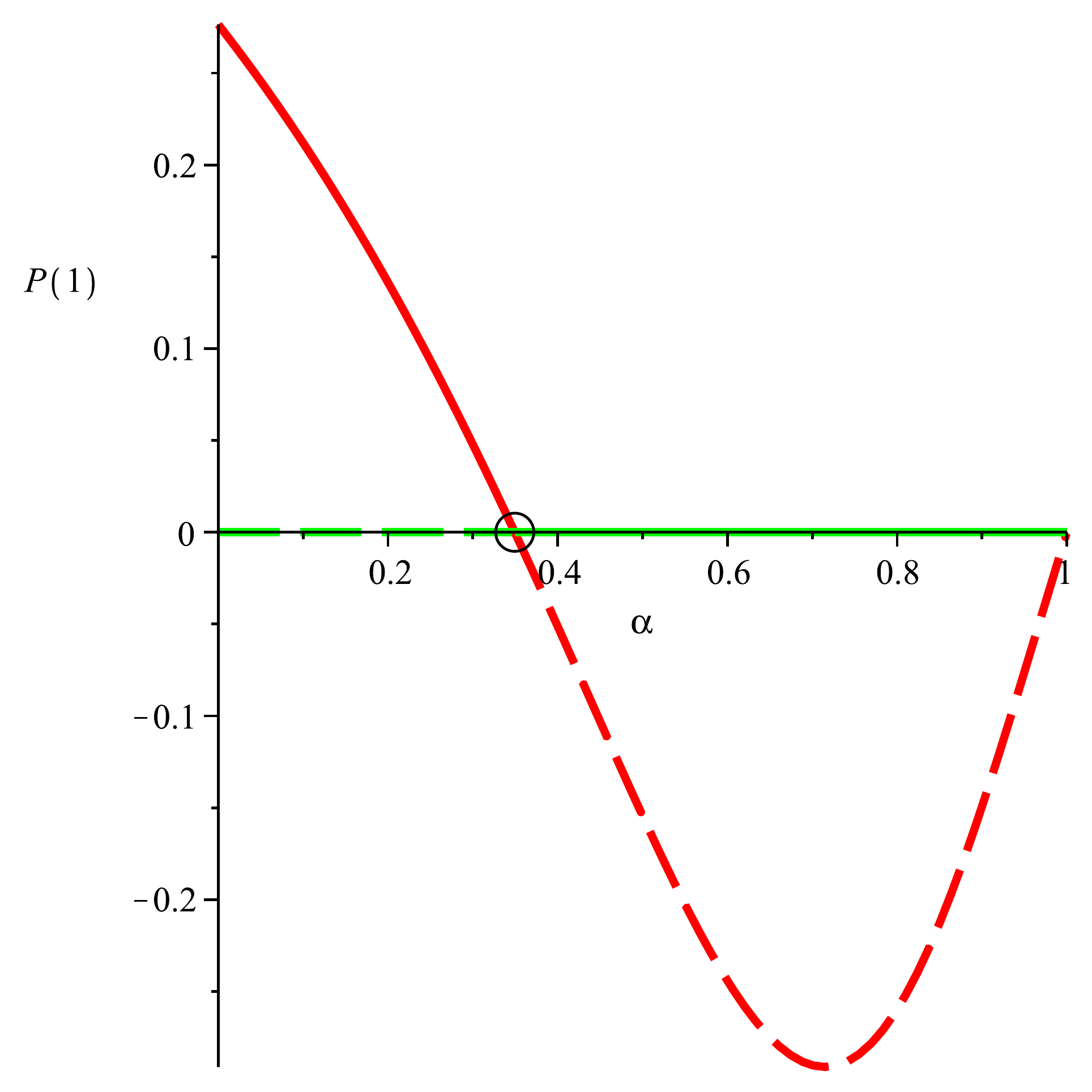} Rule 38
   \end{center}
\caption{Bifurcation diagram for local structure equations of level three for rules 6, 18, and 38.  Solid line represents stable fixed point, dashed line unstable fixed point. Transcritical bifurcation
points are circled.}\label{bif6-18-38}
\end{figure}

For $k=3$, we have 8 equations given by  (\ref{lsteq}), but again, only four of them are independent
because of consistency conditions. These four are still rather complicated, thus, to save some space
occupied by indices, we will write $\varLambda^{(3)}$ map instead of equations (\ref{lsteq}), and we will
relegate some longer expressions to the Appendix.
Assuming that denominators in (\ref{lsteq}) are positive, taking into account consistency conditions, 
and using variables $x=P(000),y=P(001), z=P(010)$ and $v=P(011)$, these four components of $\varLambda^{(3)}$ become
\begin{align}
 x&\rightarrow\frac{-d_x}{(x+y) (y+v)^2},\\
 y&\rightarrow\frac{d_y}{(x+y) (y+v)^2},\\
 z&\rightarrow\frac{d_z}{y+v},\\
 v&\rightarrow\frac{-d_v}{(x+y) (y+v)^2},
\end{align}
where $d_x$, $d_y$, $d_z$ and $d_v$ are rather complicated polynomials, defined in the Appendix.
Replacing arrows by equalities, we obtain a system of equations for fixed point. In order to solve it, it is
convenient to change variables to $X=1-x-2y-v,Y=y+v,Z=z,V=x$, or equivalently
$x = V, y = -V-Y+1-X, z = Z, v = V+2Y+X-1$. This change of variables is purposefully constructed 
in order to simplify the map, and the rationale for this choice is explained in \cite{paper50}, where
it is called ``short block representation''. In terms of block probabilities, these new variables are
\begin{align}
 X&=P(1),\\
 Y&=P(01),\\
 Z&=P(010),\\
 V&=P(000).
\end{align}
We then obtain
\begin{align}\label{r6-eqX}
 X&\rightarrow X -\alpha (2X+Y-Z+V-1) \\  \label{r6-eqY}
 Y&\rightarrow Y +\frac{Z}{Y}(X+V-1)\alpha^2 +(1-2X+Z-V)\alpha^2 +(X-2Y+Z)\alpha \\ \label{r6-eqZ}
 Z&\rightarrow Z+  {\frac {Z}{Y}}(V+X-1)(\alpha^3+\alpha)- \left(2X-Y+V-1 \right) {\alpha}^{3}\\ \nonumber
&\mbox{\hspace{5cm}} +\left(X-4 Y+3 Z \right) {\alpha}^{2}+ \left(2 Y-Z \right) \alpha\\ \label{r6-eqV}
 V&\rightarrow V+ \frac{\alpha }{(X+Y-1)^2 Y^2} d_V,                                                      
\end{align}
where
\begin{multline}
d_V=(X+Y-1) (2 X Y-X Z-Y^2+Y V-Z V-Y+Z) Y \alpha^2\\
(Y-Z)(3 X^2 Y-X^2 Z+9 X Y^2-3 X Y Z+4 X Y V-X Z V+6 Y^3-2 Y^2 Z+4 Y^2 V\\
 -Y Z V+Y V^2-6 X Y+2 X Z-9 Y^2+3 Y Z-4 Y V+Z V+3 Y-Z) \alpha \\
-(X+Y+V-1) (2 X Y-2 X Z+2 Y^2-2 Y Z+Y V-2 Y+2 Z) Y.        
\end{multline}
Fixed point of the above map can be obtained with the help of Maple symbolic solver, yielding 
\begin{align}\label{rule6-x}
X&=\frac{(1-\alpha )}{d_\alpha }
 (\alpha ^6-5 \alpha ^5+10 \alpha ^4-7 \alpha ^3-2 \alpha ^2+4 \alpha +2) 
(\alpha ^6-7 \alpha ^5+19 \alpha ^4-23 \alpha ^3+8 \alpha ^2+8 \alpha -4), \\
Y&=\frac{1-\alpha }{d_\alpha } \left( {\alpha }^{3}-2 {\alpha }^{2}+\alpha +2 \right)    \left( {\alpha 
}^{6}-7 {\alpha }^{5}+19 {\alpha }^{4}-23 {\alpha }^{3}+8 {\alpha }^{2}+8 \alpha -4 \right), \\
Z&=\frac{1-\alpha }{d_\alpha }   \left( {\alpha }^{2}-2 \alpha +2 \right)  \left( {\alpha }^{6}-7
 {\alpha }^{5}+19 {\alpha }^{4}-23 {\alpha }^{3}+8 {\alpha }^{2}+8 \alpha -4 \right), \\
V&=\frac{{\alpha }^{2}}{d_\alpha }   \left( {\alpha }^{3}-2 {\alpha }^{2}+\alpha +2 \right) \left( {\alpha }^{2}-3 \alpha +3 \right)
 \left( 2 {\alpha }^{6}-14 {\alpha }^{5}+40 {\alpha }^{4}-55 {\alpha }^{3}+28 {\alpha }^{2}+13
 \alpha -16 \right), 
\end{align}
where
\begin{equation}
 d_\alpha=2(\alpha ^{12}-11 \alpha ^{11}+54 \alpha ^{10}-152 \alpha ^9+259 \alpha ^8-248 \alpha ^7+73 \alpha ^6+95 \alpha ^5
-84 \alpha ^4-15 \alpha ^3+32 \alpha ^2+2 \alpha -8).
\end{equation}
We will call this fixed point ``active'', since it corresponds to non-zero value of $X$. 
Graph of $X$, that is, $P(1)$, as a function of $\alpha$ for this fixed point is shown in Figure~\ref{bif6-18-38}.
 It is clear that $P(1)$
 becomes negative at some point $\alpha=\alpha_c$. The value $\alpha_c$ can be obtained by solving equation $X=0$,
but unfortunately the solution is not expressible by elementary functions, thus we can only say that
$\alpha_c$ must satisfy
\begin{equation}
{\alpha_c}^6-7{\alpha_c}^5+19{\alpha_c}^4-23{\alpha_c}^3+8{\alpha_c}^2+8{\alpha_c}-4=0,
\end{equation}
yielding numerical value of $\alpha_c=0.4827758301$. Since the $X$ component of the fixed point given by eq. (\ref{rule6-x}) becomes
negative for $\alpha>\alpha_c$, and we cannot have negative probability, one can expect that the fixed point becomes unstable for $\alpha>\alpha_c$.
In order to verify this, we can use the following   general property \cite{alligood1997chaos} of maps $\mathbb{R}^n \rightarrow \mathbb{R}^n$.
In general, if $F: \mathbb{R}^n \rightarrow \mathbb{R}^n$ has a fixed point at $\mathbf{u} \in \mathbb{R}^n$,
and the Jacobian of $F$ at $\mathbf{u}$ has only eigenvalues with magnitude less than one, than $\mathbf{u}$ is stable.
If, on the other hand, magnitude of at least one of the eigenvalues of the Jacobian is greater than one, than 
$\mathbf{u}$ is unstable.

\begin{figure}
 \begin{center}
 \includegraphics[scale=0.35]{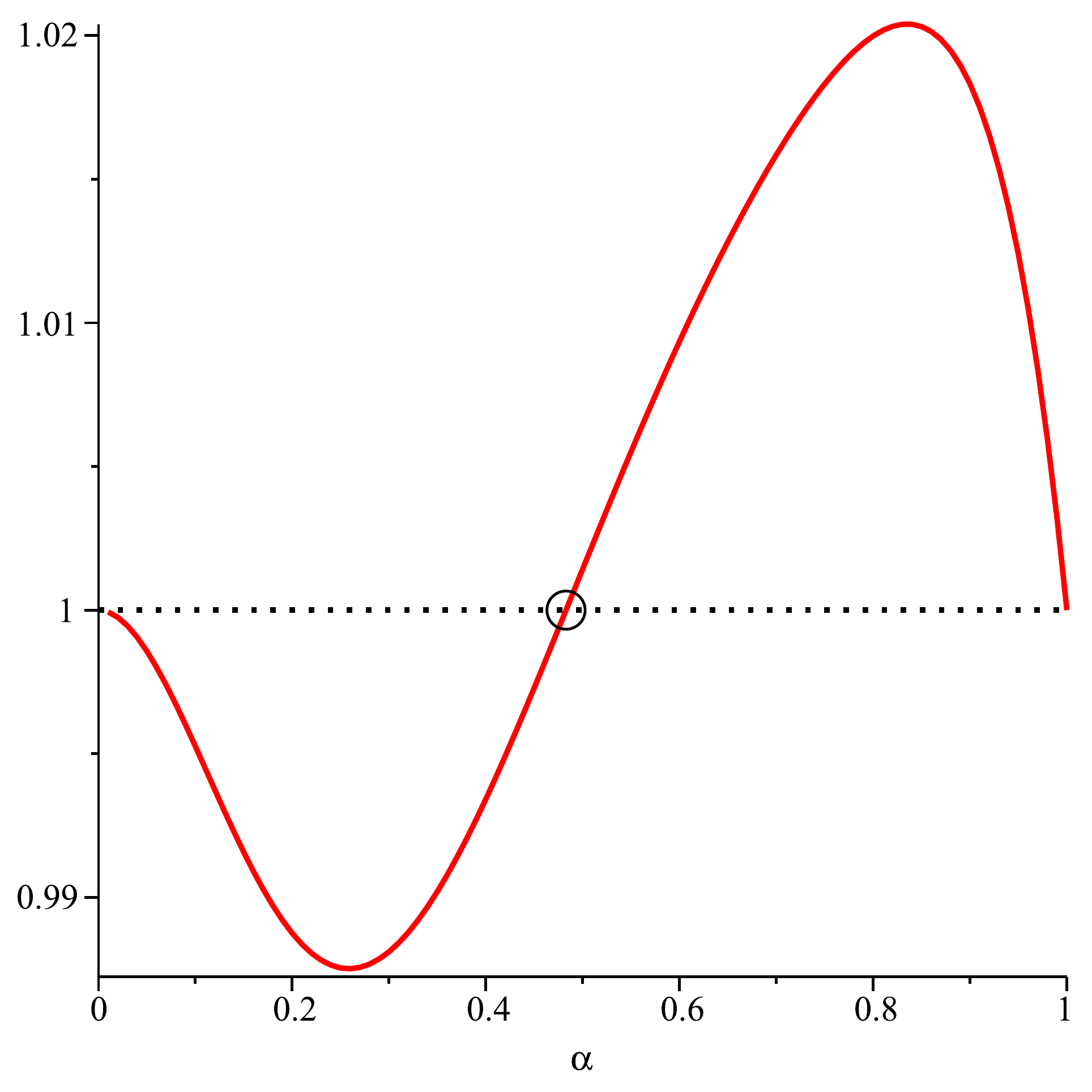}\includegraphics[scale=0.35]{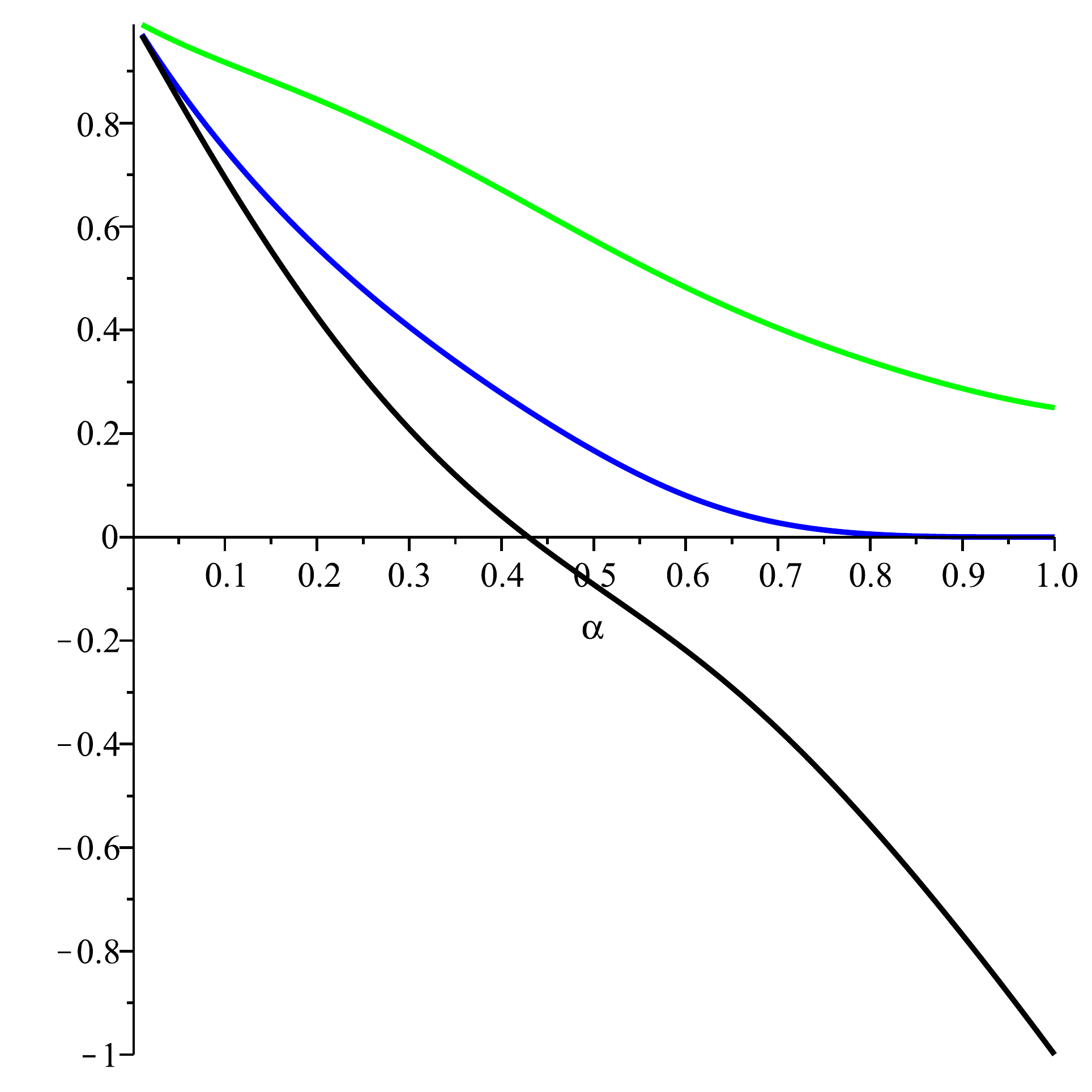}
   \end{center}
\caption{Graphs of eigenvalues of the Jacobian for rule 6. The eigenvalue which becomes larger
than 1 for $\alpha>\alpha_c$ is shown on the left, while the remaining
three eigenvalues on the right.}\label{r6eigenvalues}
\end{figure}

For the map $\mathbb{R}^4 \rightarrow \mathbb{R}^4$ 
given by eqs. (\ref{r6-eqX}--\ref{r6-eqV}), all four eigenvalues of the Jacobian
evaluated at the fixed point given by eq. (\ref{rule6-x}) are real. Using symbolic algebra
software, it is possible to obtain
explicit expressions for these eigenvalues as a function of $\alpha$, but they
are too complicated to be included here. Instead, we include their graphs, shown
in Figure~\ref{r6eigenvalues}. 
One can clearly see that magnitudes of three of these eigenvalues remain
in the interval $[0,1]$, while the largest one becomes greater than one when
$\alpha$ is sufficiently large. This happens when $\alpha>\alpha_c$, therefore we can
conclude that the fixed point becomes unstable when $\alpha>\alpha_c$.

Remembering our ``thick bar'' convention, one can verify that there exists 
another fixed point of eqs. (\ref{lsteq}) for $k=3$. In terms of our new variables, it is given by
 $X=0,Y=0,Z=0,V=1$. We will call this fixed point ``absorbing''.
Its stability  cannot be established by the same method, because the mapping
given by eqs. (\ref{lsteq}) is not differentiable at this point. Nevertheless, it can be determined numerically
by simply iterating the map and checking if it converges to the fixed point or not. Using this method, we verified that
the fixed point $X=0,Y=0,Z=0,V=1$ is stable  when $\alpha>\alpha_c$ and unstable for  $\alpha<\alpha_c$. This means
that at $\alpha=\alpha_c$ a transcritical bifurcation takes place, with exchange of stability between fixed points.
This is illustrated in Figure~\ref{bif6-18-38}, which shows bifurcation diagram for the $X$ component.
We can therefore summarize our results by saying that the local structure approximation
``predicts'' that  $P_t(1)$ behaves
as follows
\begin{equation}
 \lim_{t \to \infty} P_t(1)=\begin{cases}
                             \frac{(1-\alpha )(\alpha ^6-5 \alpha ^5+10 \alpha ^4-7 \alpha ^3-2 \alpha ^2+4 \alpha +2) 
(\alpha ^6-7 \alpha ^5+19 \alpha ^4-23 \alpha ^3+8 \alpha ^2+8 \alpha -4)}{2(\alpha ^{12}-11 \alpha ^{11}+54 \alpha ^{10}-152 \alpha ^9+259 \alpha ^8-248 \alpha ^7+73 \alpha ^6+95 \alpha ^5
-84 \alpha ^4-15 \alpha ^3+32 \alpha ^2+2 \alpha -8)} 
  & \text{if $\alpha<\alpha_c$},\\
                             \mbox{\hspace{6cm}} 0 & \text{if $\alpha \geq \alpha_c$}.
                            \end{cases}
\end{equation}
Taylor expansion of the above expression for $\alpha<\alpha_c$ yields
\begin{equation}
\lim_{t \to \infty} P_t(1)=-A(\alpha-\alpha_c) +
O((\alpha-\alpha_c)^2)
\end{equation}
where $A=30 {\alpha_c}^5-272 {\alpha_c}^4+917 {\alpha_c}^3-1390 {\alpha_c}^2+875 {\alpha_c}-166$,
which means that the local structure theory predicts, as expected, $\lim_{t \to \infty} P_t(1) \sim (\alpha-\alpha_c)^1$,
i.e., critical exponent $\beta=1$.

In conclusion,   the local structure approximation at level three correctly
predicts the existence of the phase transition in rule 6, and, moreover, it correctly predicts the direction of the transition:
the active phase appears as $\alpha$ decreases. The critical value of $\alpha_c=0.4827758301$, however, is very far
from the experimentally determined value reported in \cite{fates:inria-00138051}, which is $\alpha_c=0.2825$.
\section{Local structure approximation for $\alpha$-asynchronous rules 18, 50, and  134}
\begin{figure}
 \begin{center}
Rule 50 \hspace{8cm} Rule 106\\
\includegraphics[scale=0.35]{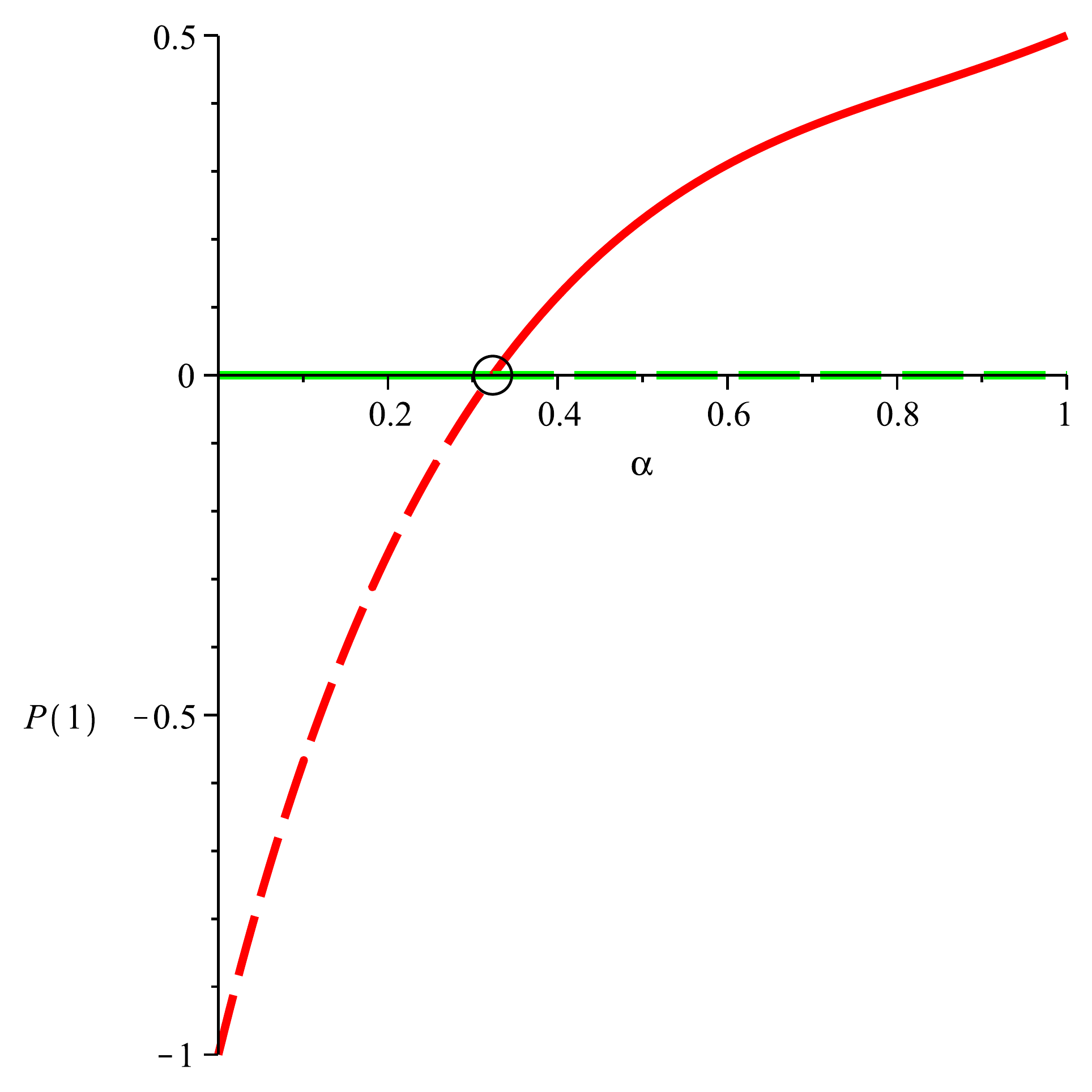} \includegraphics[scale=0.35]{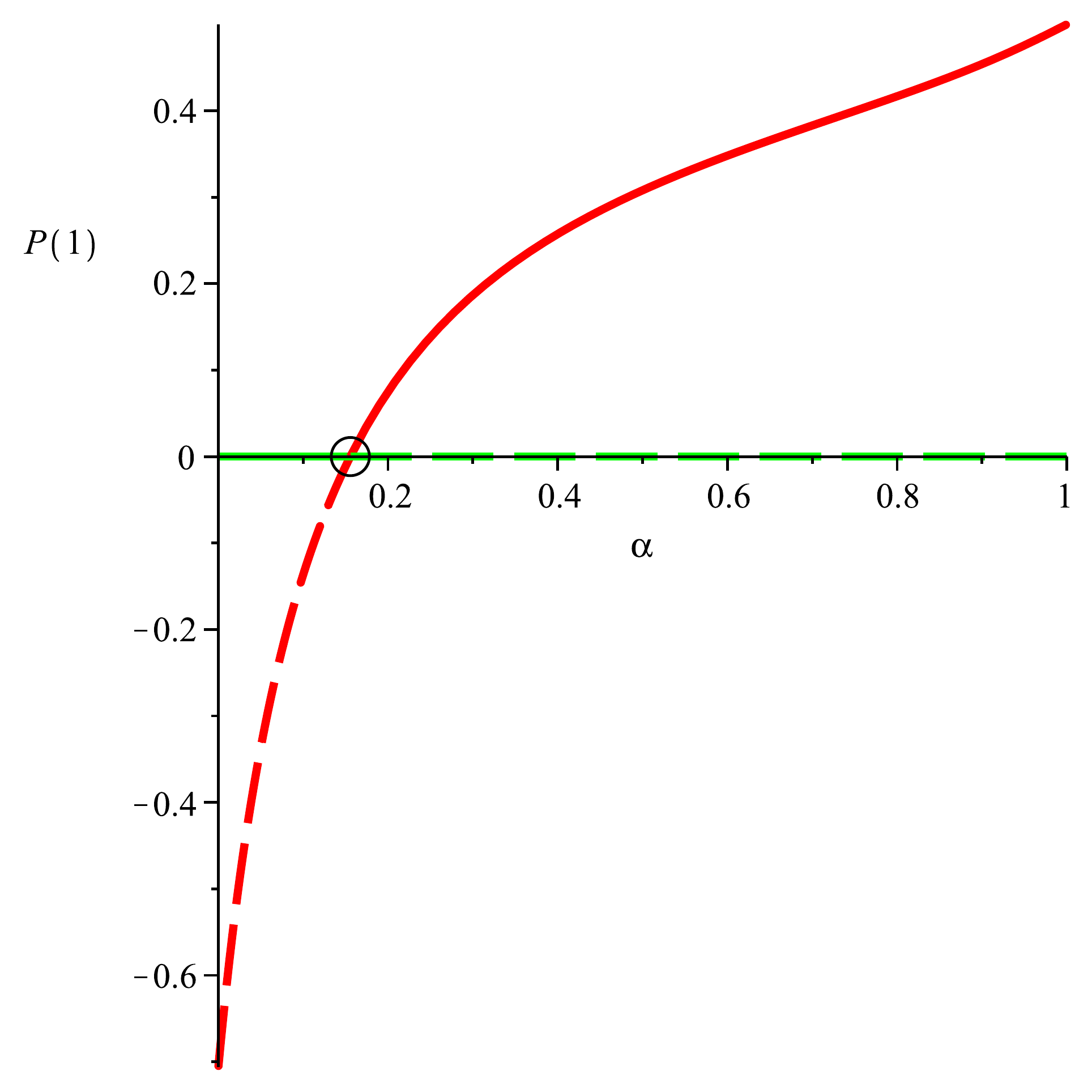}\\
 \includegraphics[scale=0.35]{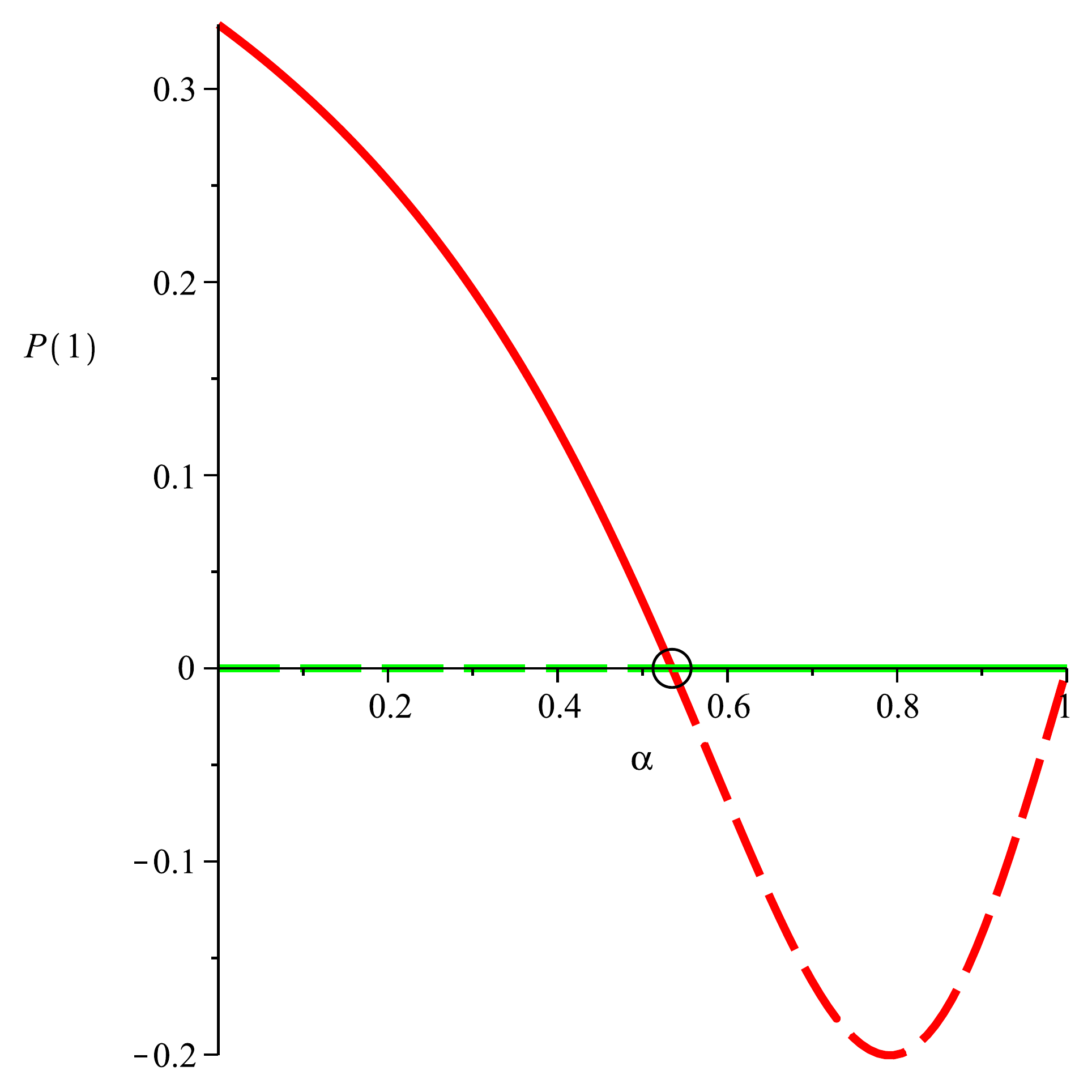} Rule 134
   \end{center}
\caption{Bifurcation diagram for local structure equations of level three for rules 50, 106, and 134.  Solid line represents stable fixed point, dashed line unstable fixed point. Transcritical bifurcation
points are circled.}\label{bif50-106-134}
\end{figure}

For rules 18, 50, and 134, the absorbing fixed point of the local structure map $\varLambda^{(3)}$ is the same as before, $X=0,Y=0,Z=0,V=1$.
Active fixed point of $\varLambda^{(3)}$ can also be found
using procedure outlined in the previous section, with the help of Maple symbolic algebra software.
We only give expressions for the $X$ component of the fixed point, that is, for $P(1)$. 
\begin{itemize}
\item Rule 18
\begin{multline}\label{r18dens}
P(1)=\frac{-20  {\alpha}^5+156  {\alpha}^4-473  {\alpha}^3+670  {\alpha}^2-304  {\alpha}+\sqrt[3]{A_1}-96}
{24  {\alpha}^4-120  {\alpha}^3+270  {\alpha}^2-96  	{\alpha}-192} \\
+
\frac{A_2}
{\sqrt[3]{A_1} (24  {\alpha}^4-120  {\alpha}^3+270  {\alpha}^2-96  {\alpha}-192)},
\end{multline}
where
$A_1$ and $A_2$ are defined in the Appendix.
The critical value $\alpha_c$ satisfies fourth order equation
\begin{equation}
 \alpha_c^4-7 \alpha_c^3+16 \alpha_c^2-16 {\alpha_c}+4=0.
\end{equation}
Although it is possible to express $\alpha_c$ in terms of radicals, the expression is rather long,
thus we only give its numerical value here, $\alpha_c=0.3605426781$.

\item Rule 50
\begin{multline}
 P(1)= 1-{\frac { ( 2 {\alpha }^{2}-4 \alpha +2 ) \sqrt{(1-\alpha )(16 \alpha ^4-57 \alpha ^3+95 \alpha ^2-75 \alpha +25)}  }{ 2 ( 1-2 {\alpha }^{2}
 )  ( 2 {\alpha }^{2}-6 \alpha +5 ) }}\\
+ {\frac {8 {\alpha }^{5}-34 {\alpha }^{4}+58 {\alpha }^{3}-56 {\alpha }^{2}+35 \alpha -10}
{ 2 ( 1-2 {\alpha }^{2}
 )  ( 2 {\alpha }^{2}-6 \alpha +5 ) }}
\end{multline}

Bifurcation takes place at
\begin{equation}
\alpha_c=\frac{5}{6}-\frac{1}{12}\,\left(188+12\,\sqrt {249}\right)^{1/3}+{\frac {2}{3}{\left({188+12
\,\sqrt {249}}\right)^{-1/3}}} \approx 0.3233950179.
\end{equation}

\item Rule 134
 \begin{equation}
P(1)= 1-2\,{\frac {115\,{\alpha }^{3}-298\,{\alpha }^{2}+280\,\alpha -90+3 \left(5\alpha -4\right) \sqrt {8\,
{\alpha }^{2}-16\,\alpha +9 }}{575\,{\alpha }^{3}-
1430\,{\alpha }^{2}+1253\,\alpha -378}}.
\end{equation}
The transcritical bifurcation occurs at $\alpha={\frac {31-\sqrt{41}}{46}}\approx 0.5347146906$.
\end{itemize}
In all three cases, we have exchange of stability of active and absorbing fixed point at $\alpha=\alpha_c$.
Resulting bifurcation diagrams for rule 18 is shown in Figure \ref{bif6-18-38}, while
diagrams for rule 50  and  134  in Figure \ref{bif50-106-134}.
\section{Remaining rules}
For rule 38, the active fixed point of $\varLambda^{(3)}$ can be computed, but the expression is very long.
 The critical point $\alpha_c=0.3493360869$ is a solution of 
\begin{equation}
 8 \alpha_c^7-42 \alpha_c^6+106 \alpha_c^5-155 \alpha_c^4+142 \alpha_c^3-82 \alpha_c^2+28 \alpha_c-4=0.
\end{equation}

For rule 106, active fixed point of $\varLambda^{(3)}$ can also be computed, but, similarly as in the case of rule 38, the resulting
formulas have hundreds of terms, thus we omit them here.
 The critical point $\alpha_c=0.1556040146$ is a solution of 
\begin{equation}
 18 \alpha_c^7-131 \alpha_c^6+464 \alpha_c^5-1512 \alpha_c^4+2678 \alpha_c^3-2160 \alpha_c^2+688 \alpha_c-64=0.
\end{equation}
Absorbing fixed points for rules 38 and 106 are the same as before, and $X=0,Y=0,Z=0,V=1$.
Bifurcation diagrams of these rules are shown in Figures \ref{bif6-18-38} and \ref{bif50-106-134}.

\begin{figure}
 \begin{center}
Rule 26 \hspace{8cm} Rule 146\\
\includegraphics[scale=0.35]{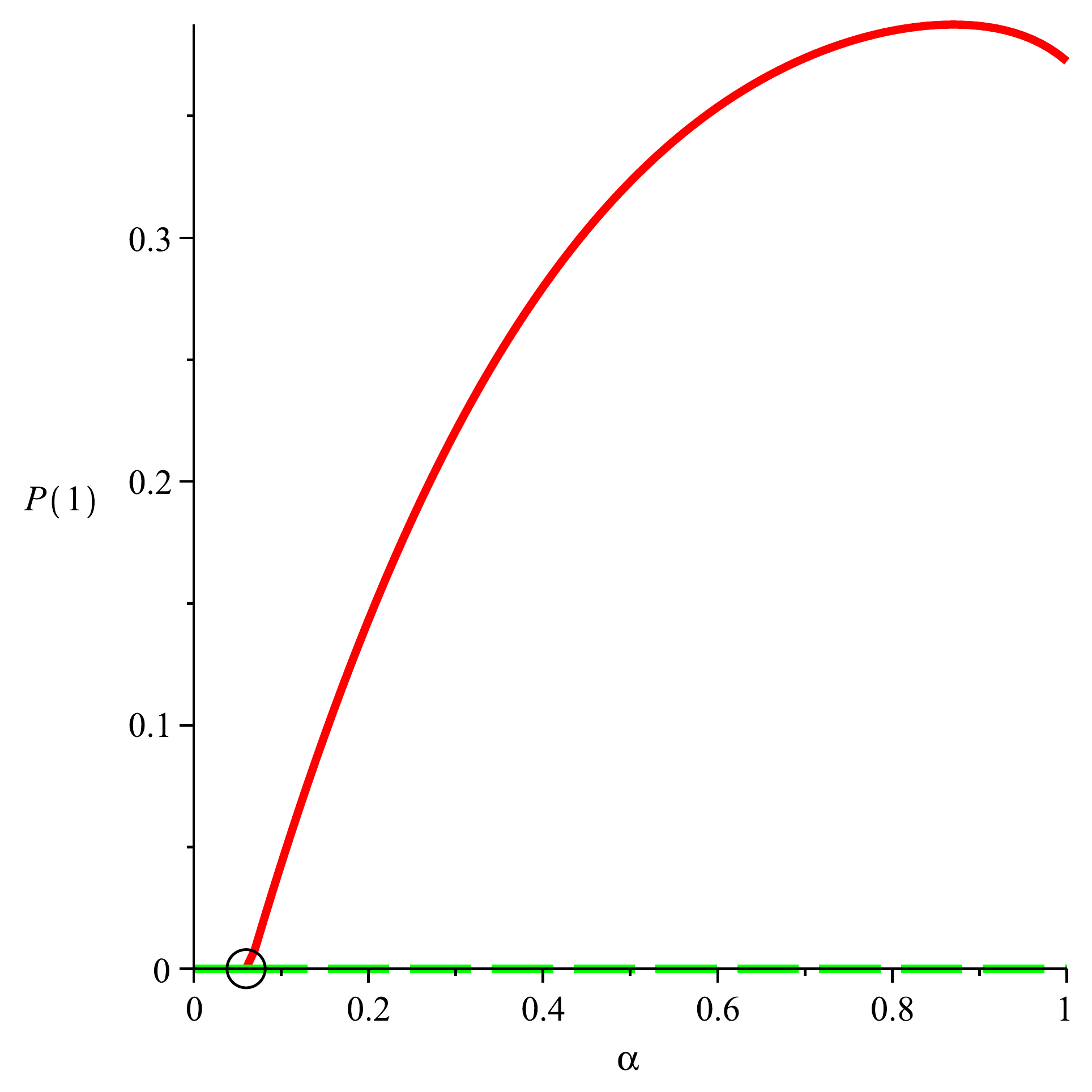} \includegraphics[scale=0.35]{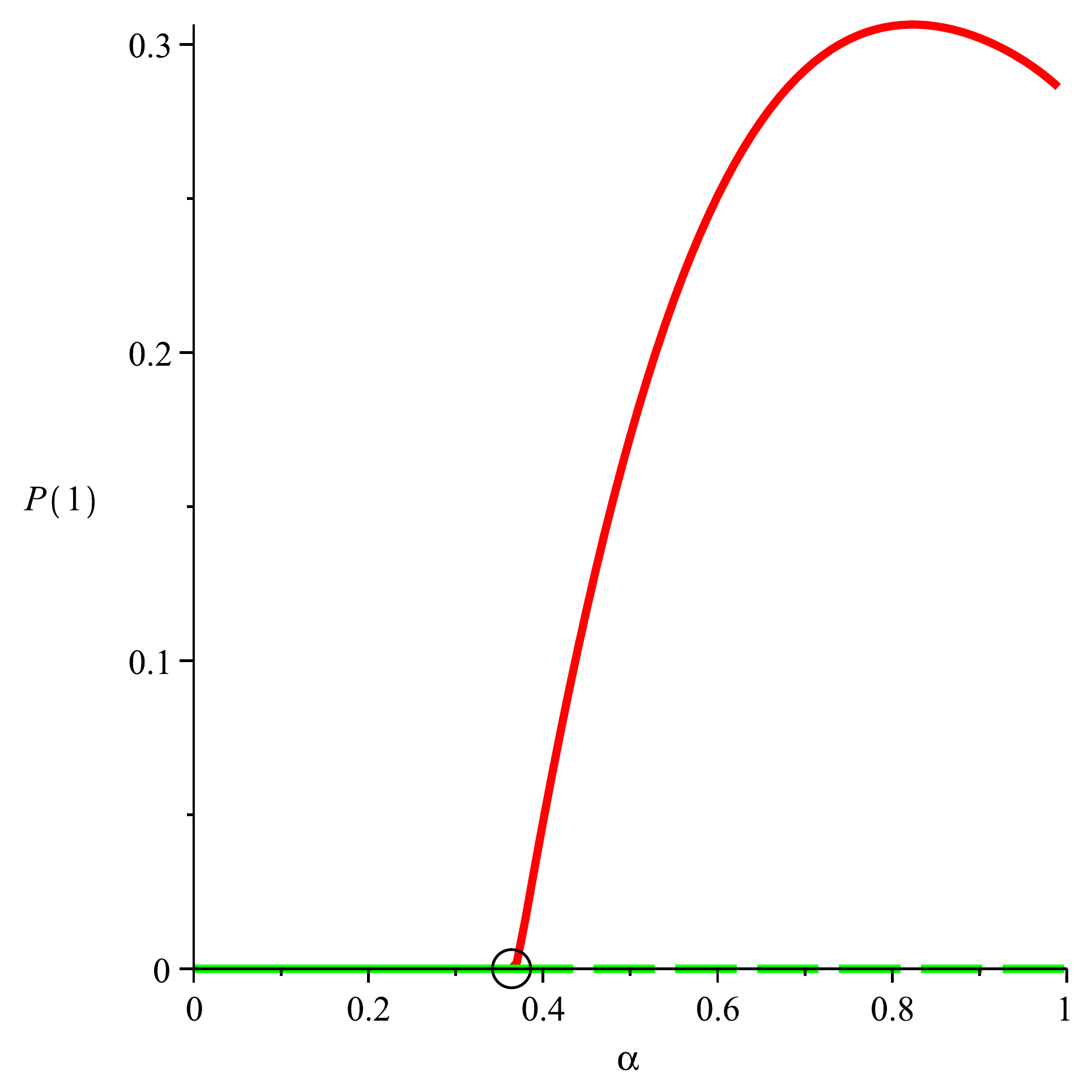}\\
 \includegraphics[scale=0.35]{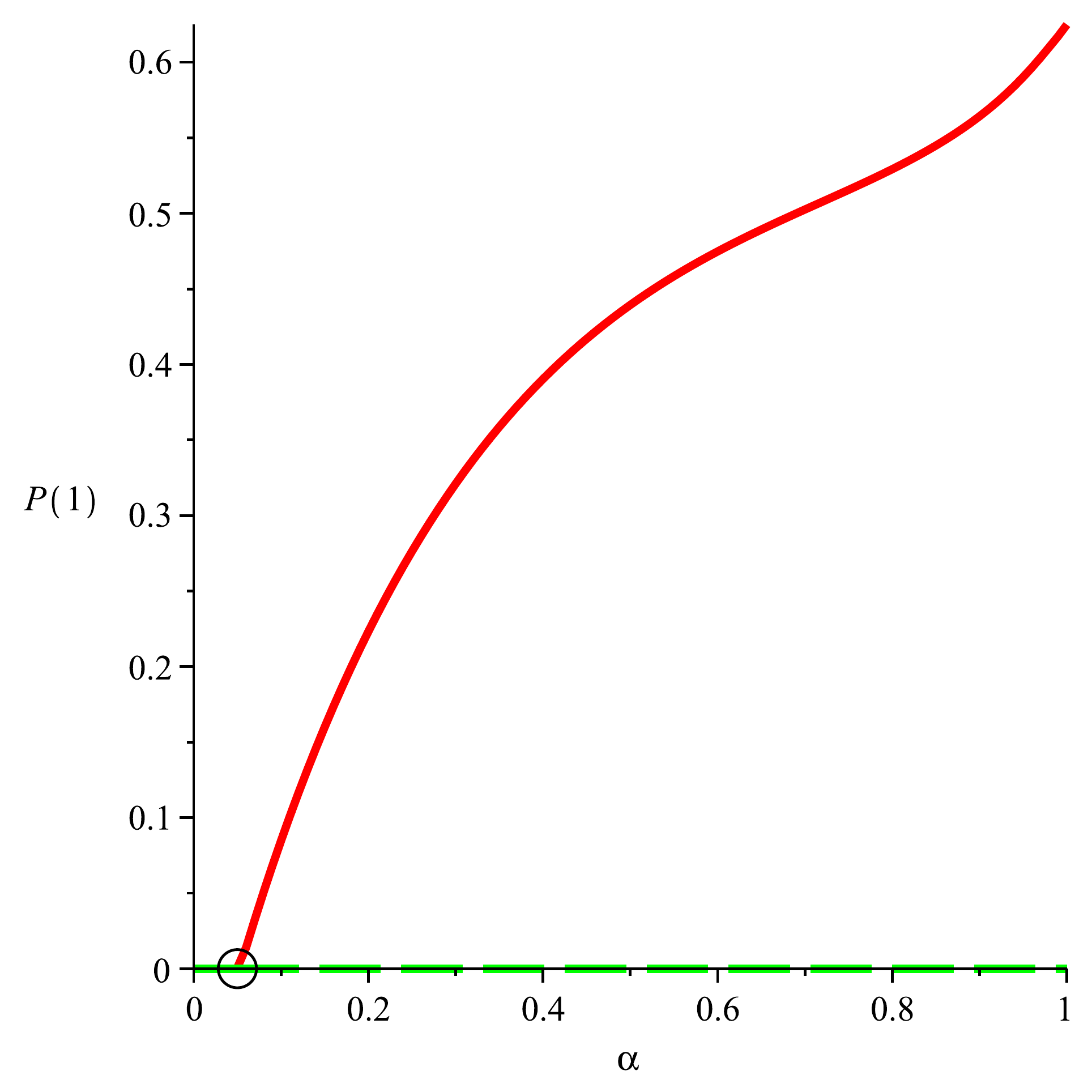} Rule 58
   \end{center}
\caption{Partial bifurcation diagrams for local structure equations of level 4 rules 26 and 146,
and level 5 for rule 58. Diagrams were
obtained numerically with $10^5$ iterations. Solid line represents stable fixed point, dashed line unstable fixed point. Transcritical bifurcation
points are circled.}\label{bif26-146-58}
\end{figure}

For the three remaining rules, 26, 58, and 146, the local structure map $\varLambda^{(3)}$ does not exhibit
a transcritical bifurcation, so it is necessary to consider higher order maps, of level four (for rules 26 and 146)
and five (for rule 58). Absorbing fixed points of these maps have the same structure as previously described,
with $X=0$. Unfortunately, equations for their active fixed points cannot be
solved even with the help of symbolic algebra software, due to the size of relevant equations. It is, however, possible to 
find the stable branch of the bifurcation diagram by iterating these maps many times, so they converge sufficiently close to the stable fixed point.
We performed such iterations for all three cases, and the results are shown in Figure \ref{bif26-146-58}. Even though the
unstable branch of the active fixed point is missing, it is evident that the active phase appears abruptly as $\alpha$
increases, which provides a strong evidence for transcritical bifurcation.

\section{Higher order local structure maps}
As we could see in previous sections, local structure approximation of order 3 to 5 can predict existence
of the phase transition for all DP rules. The local structure map for each of these rules exhibits a transcritical bifurcations,
and the direction of the bifurcation agrees with the direction of the phase transition observed experimentally, that
is, if the active phase appears (disappears) as $\alpha$ increases, then the non-zero fixed point of the local structure map
 becomes stable (unstable) as $\alpha$ increases. The point at which the transcritical bifurcation occurs is,
however, rather far from the critical point observed experimentally.  Thus one could say that the 
local structure approximation of order 3 to 5  approximates  the \emph{value} of $\alpha_c$ quite poorly.

Can this be improved by increasing the order of the local structure approximation? The answer is indeed yes,
although we cannot expect to be able to find explicit symbolic expressions for fixed points of eq. (\ref{lsteq}) when $k$ is large. 
One can, however, iterate $\varLambda^{(k)}$ many times, starting from some generic initial condition, and when
this is done, the orbit of $\varLambda^{(k)}$ indeed converges to a stable fixed point, which, depending on the value of
$\alpha$, can be zero or non-zero.

\begin{figure}
 \begin{center}
   \includegraphics[scale=0.49]{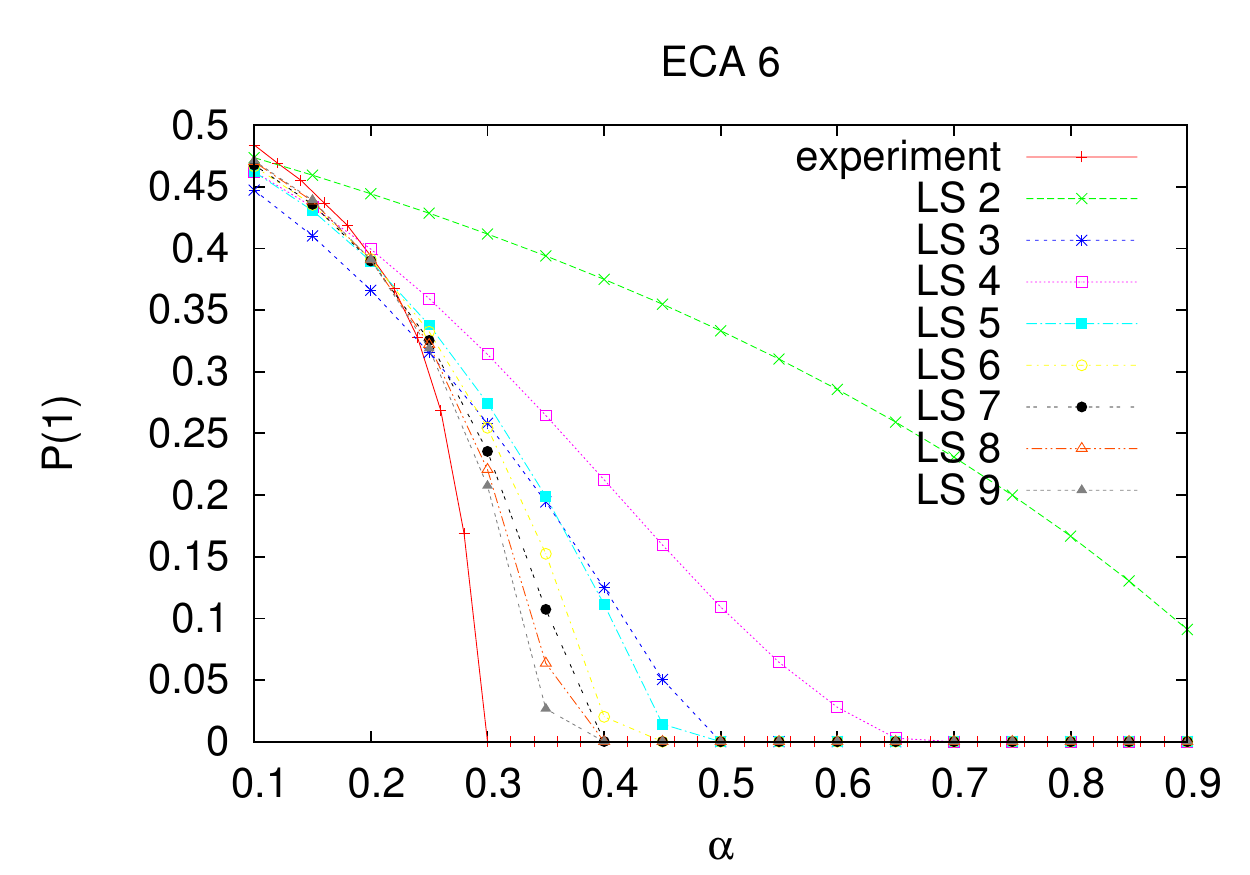}\includegraphics[scale=0.49]{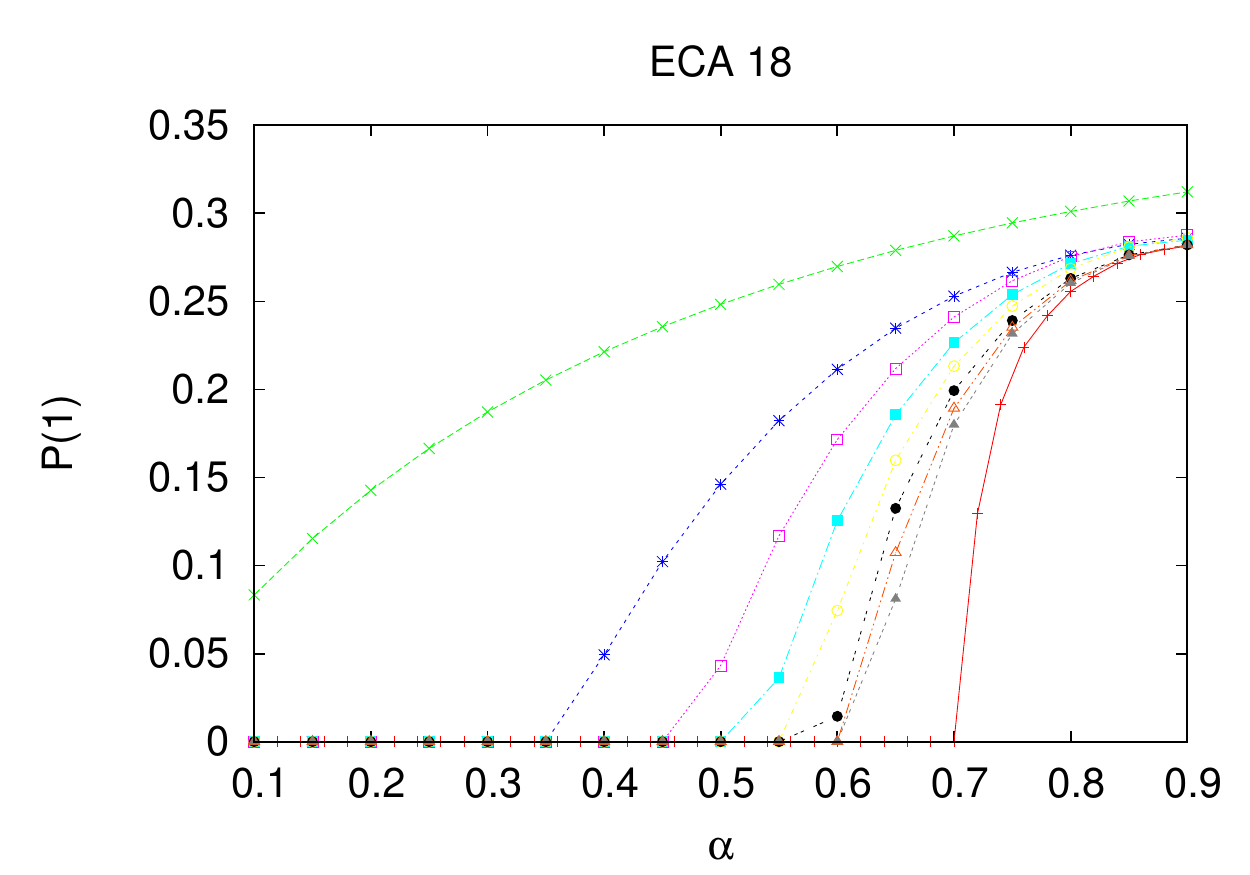}\\
   \includegraphics[scale=0.49]{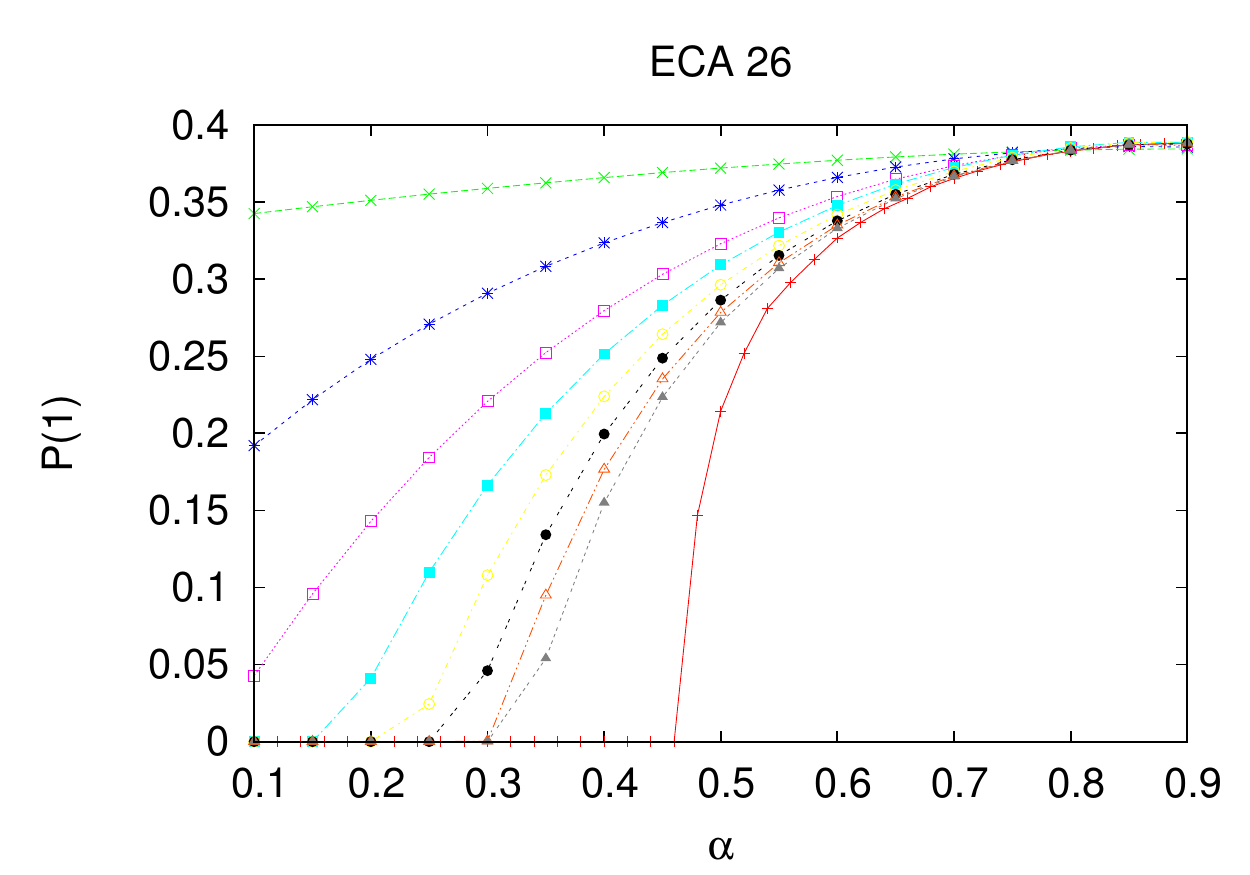}\includegraphics[scale=0.49]{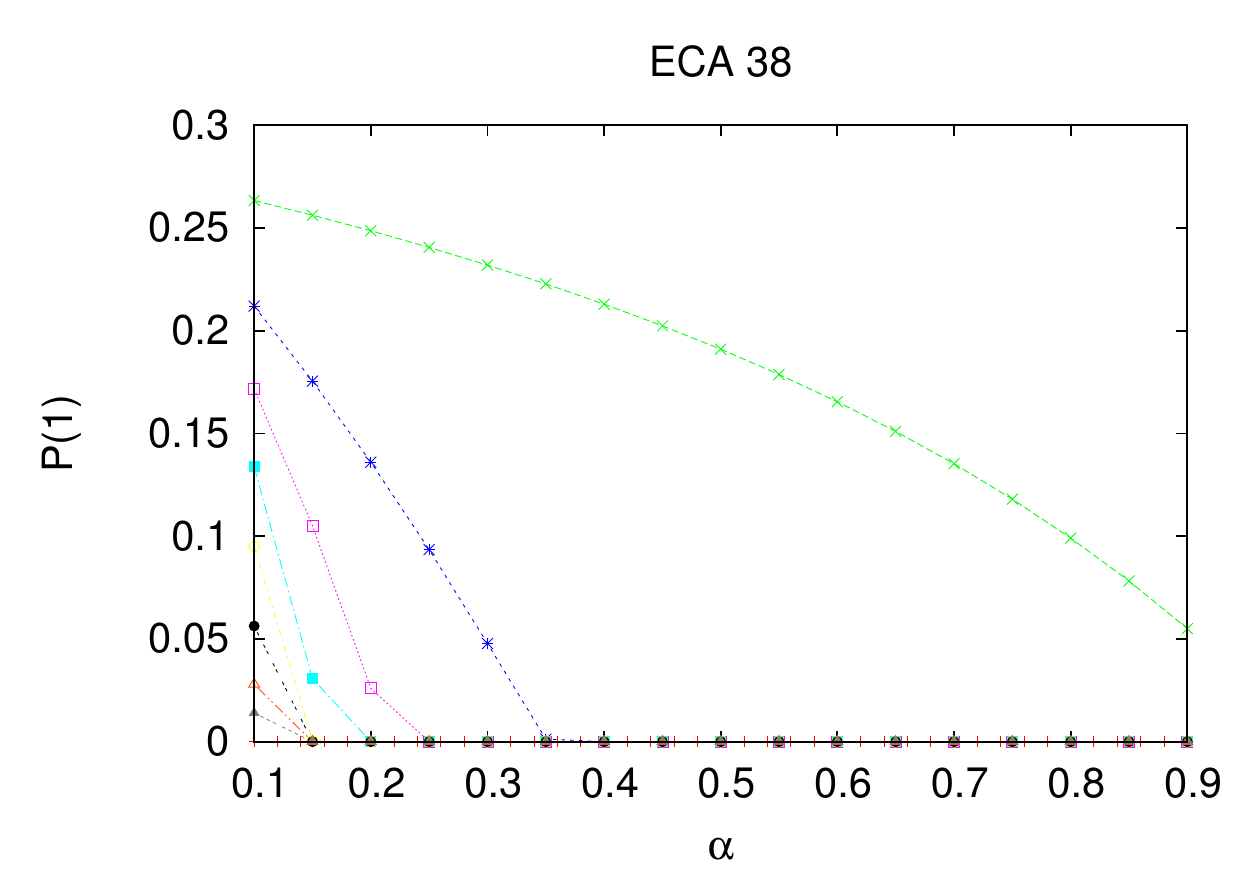}\\
   \includegraphics[scale=0.49]{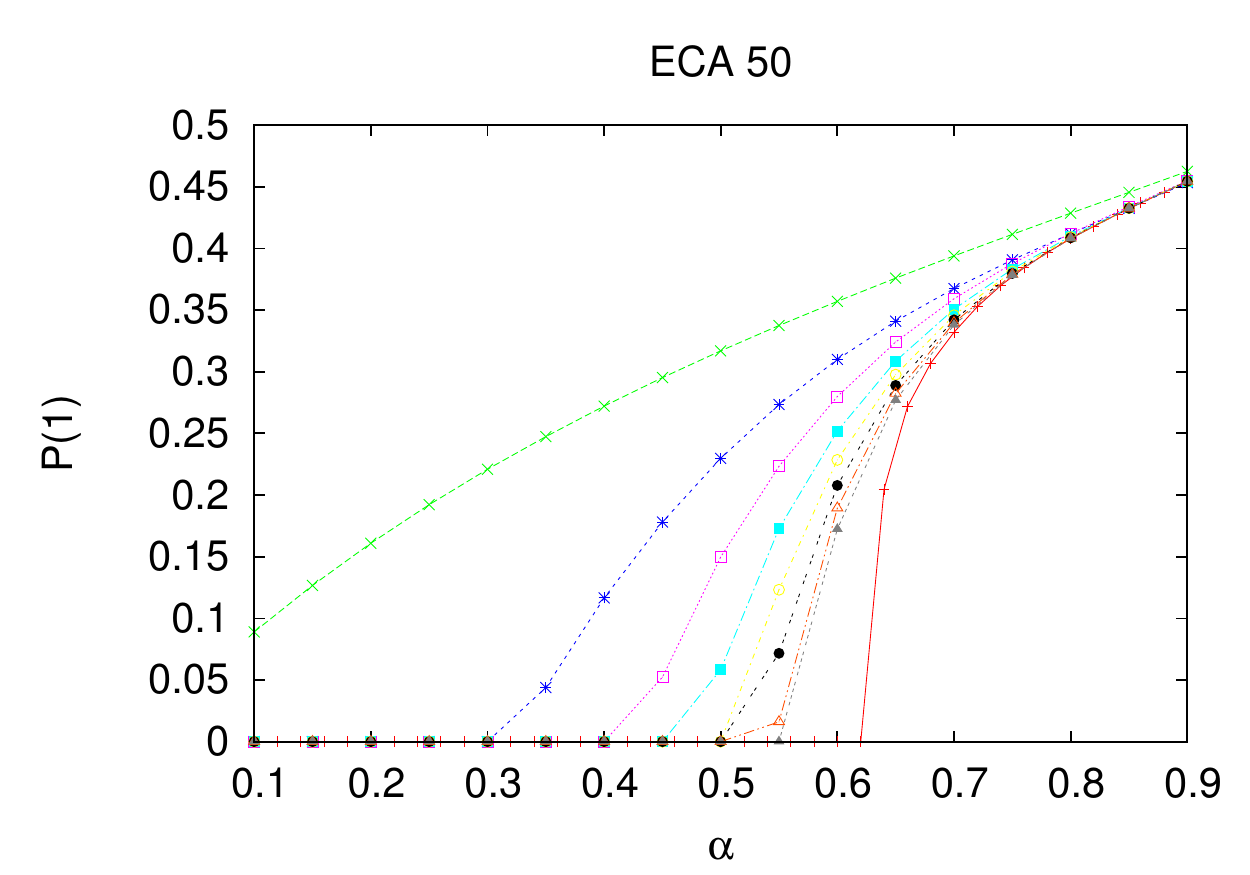}\includegraphics[scale=0.49]{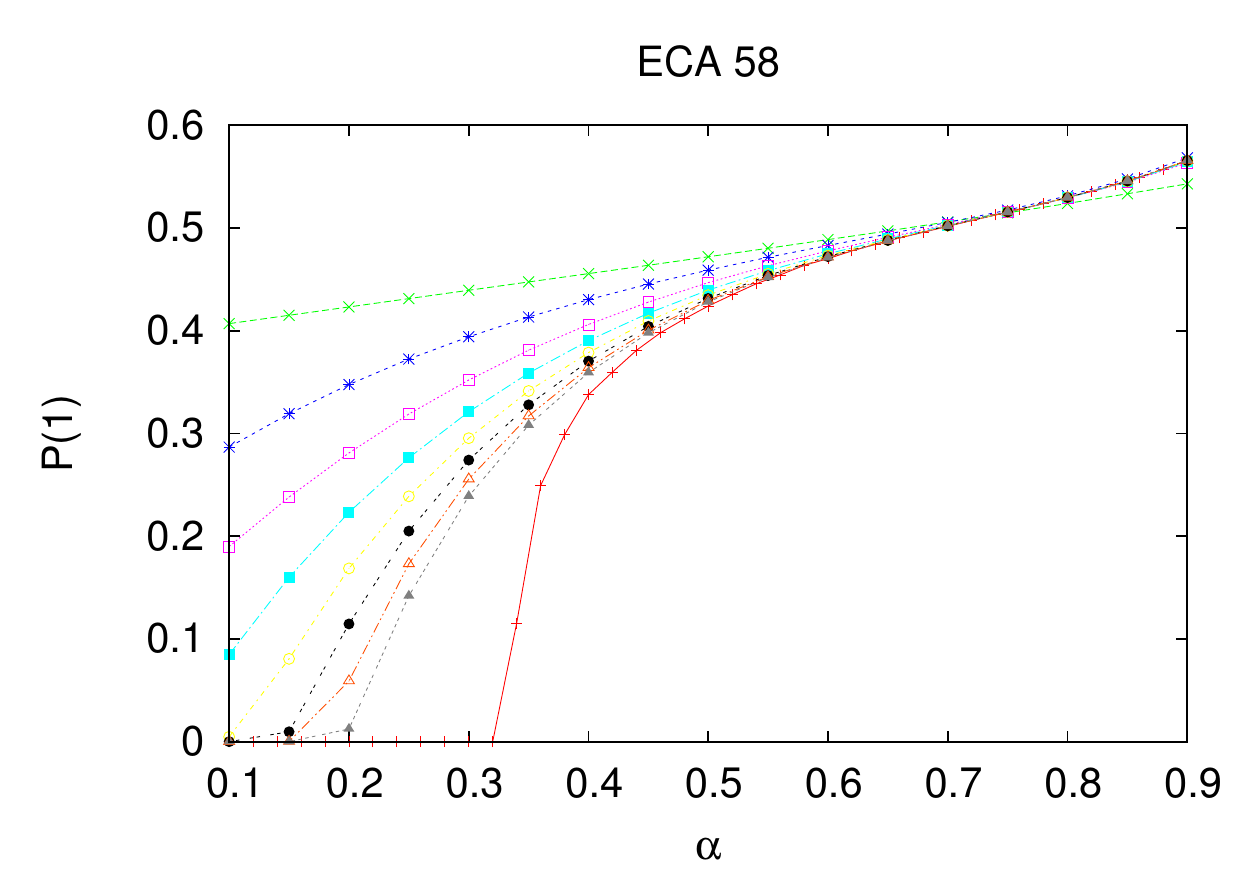}\\
   \includegraphics[scale=0.49]{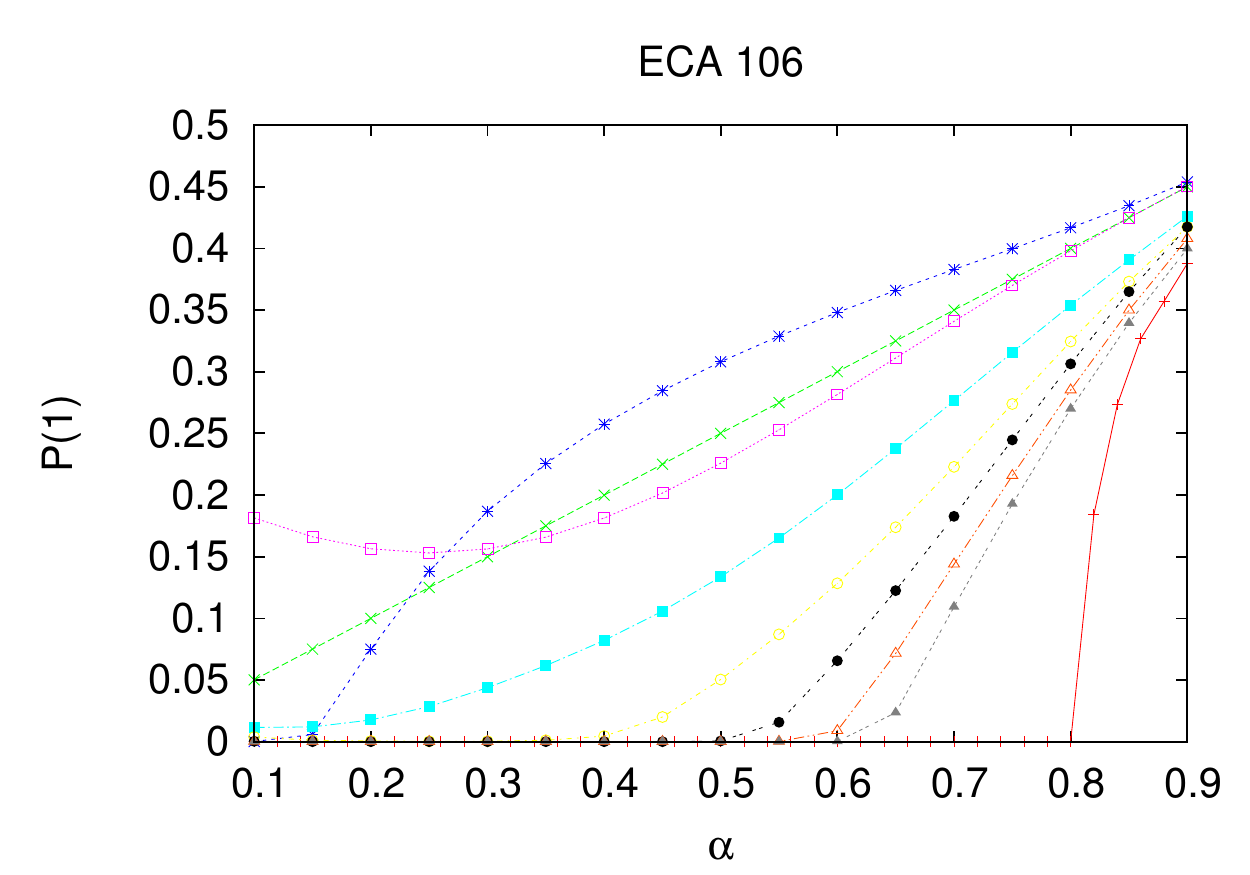}\includegraphics[scale=0.49]{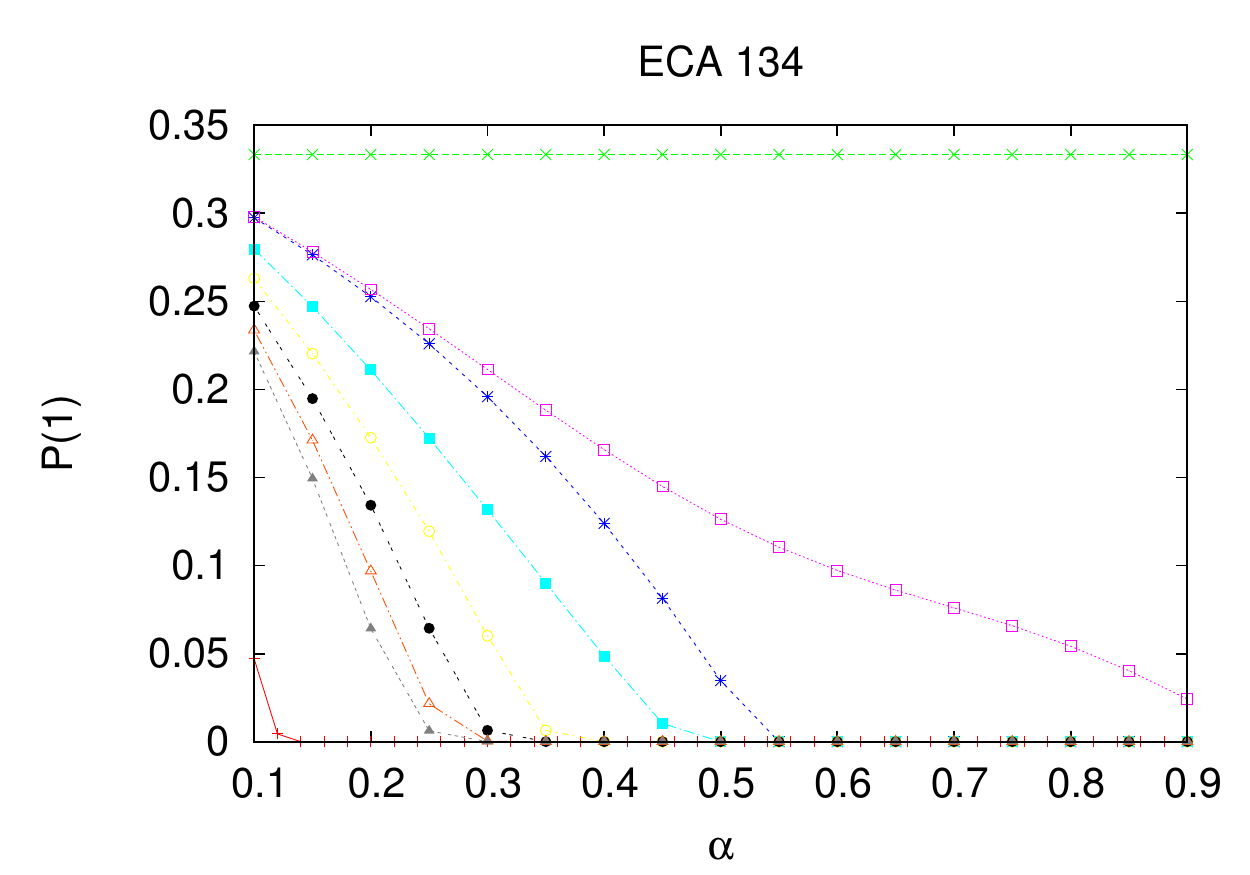}\\
   \includegraphics[scale=0.49]{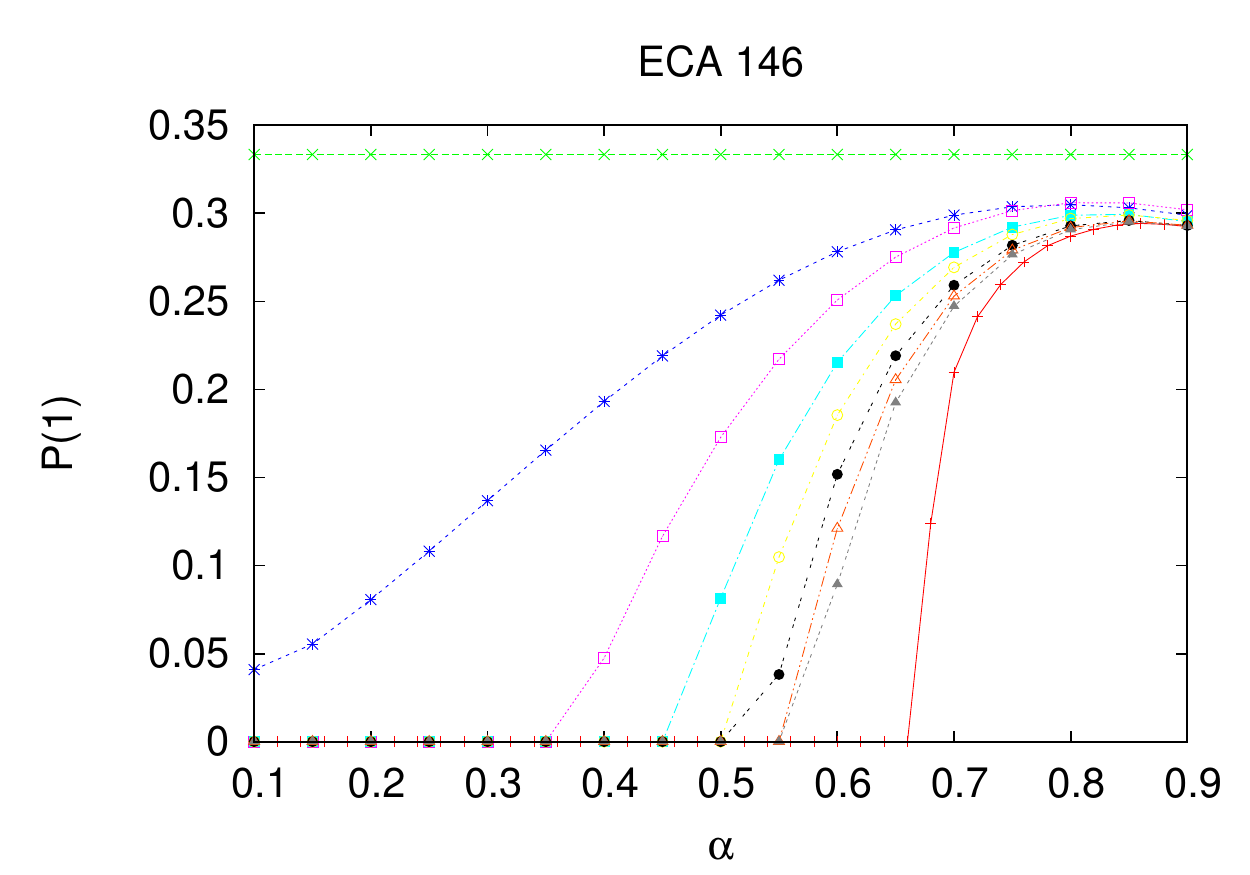}%
   \end{center}
\caption{Experimental results together with local structure approximations up to level 9.}\label{experiments}
\end{figure}
We performed iterations of $\varLambda^{(k)}$ maps for $k=2 \ldots 9$ for all DP rules, and plotted 
$P_t(1)$ as a function of $\alpha$ after $t=10^4$ iterations. Results are shown in Figure~\ref{experiments}, together
with curves obtained ``experimentally'' by iterating a given rule for $10^5$ steps, 
using randomly generated initial configurations with $4\cdot 10^4$ sites and periodic boundary
conditions. Once can clearly see that local structure maps  not only predict existence of phase
transitions, but also seem to approximate behaviour of density curves with increasing accuracy as the order of
local structure approximation increases.
%
%



\section{Conclusions}
We have demonstrated that the local structure approximation of sufficiently high level 
correctly predicts existence of phase transitions in $\alpha$-asynchronous rules belonging 
to DP universality class. The phase transition manifests itself in local structure maps as
transcritical bifurcation. The direction of the transition is predicted correctly by the local structure theory, and,
more importantly, the location of the bifurcation point approximates better and
better the location of the phase transition point as the level of local structure approximation
increases. 

Based on the evidence presented in this paper, we suspect that the same may be true for other probabilistic 
CA rules belonging to DP universality class. As mentioned in the introduction,
it already known to be  true for the probabilistic mixture of rules 182 and 200 studied by Mendon\c{c}a  and de Oliveira
\cite{Mendoca2011}. We plan to investigate this conjecture for other PCA rules.

At the same time, it seems that for rules which exhibit a phase transition, but \emph{do not} belong
to DP universality class, the local structure theory does not seem to predict existence of the
phase transition at all. We investigated the only rule of this type among $\alpha$-asynchronous ECA,
namely $\alpha$-asynchronous rule 178,
belonging, according to~\cite{FatJCA09}, to $\mathrm{DP}_2$ universality class.
We found that up to level 9, local structure maps for this rule do not exhibit any bifurcations.

The question why does the local structure predict existence of phase transitions in DP class, but fails
for rules outside of this class, is obviously the most interesting one. While we were not able
to answer this question so far,  we might offer some plausible speculations. At the heart of the local structure
approximation is the Bayesian extension, which can also be understood as maximal entropy
approximation \cite{paper50}. It is, therefore, reasonable to assume that for rules which
produce somewhat ``disordered''  configurations, the local structure approximation may work well,
whereas for rules exhibiting more ``order'' (or strongly pronounced spatio-temporal features), the
approximation may be less accurate. Whether it is possible to express this conjecture in a more
formal language, it remains to be seen.
\section{Acknowledgments}
H. Fuk\'s acknowledges financial support from the Natural Sciences and
Engineering Research Council of Canada (NSERC) in the form of Discovery Grant.
This work was made possible by the facilities of the Shared
Hierarchical Academic Research Computing Network (SHARCNET:www.sharcnet.ca) and
Compute/Calcul Canada.

\section{Appendix}
Definitions of polynomials $d_x$, $d_y$, $d_z$ and $d_v$ for local structure equations for rule 6:
\begin{multline}
d_x= ( x+y )  ( v+y )  ( 3 {v}^{2}+vx-zv-v+8 
yv+yx-y-2 zy+5 {y}^{2} ) {\alpha }^{3}\\
- ( v+y-z ) 
 ( 3 {v}^{2}x+3 y{v}^{2}-vxz+3 {y}^{2}v-zvy+4 yvx+x{y}^{2}
 ) {\alpha }^{2}\\
- ( v+y ) y ( 2 yv+vx+2 {y}^{2}-2 z
y+yx-2 xz ) \alpha -x ( v+y ) ^{2} ( x+y ) 
\end{multline}
\begin{multline}
d_y= ( x+y )  ( v+y )  ( 3 {v}^{2}+vx-zv-v+8 
yv+yx-y-2 zy+5 {y}^{2} ) {\alpha }^{3}\\
+ ( 6 zvxy-17 vx{y}^{2}
-16 {v}^{2}xy+2 yvx+4 {v}^{2}xz+4 z{v}^{2}y+2 zx{y}^{2}-2 v{x}^{
2}y+5 {y}^{2}zv-{z}^{2}vx-yv{z}^{2}\\
-13 {y}^{3}v-14 {v}^{2}{y}^{2}-6
 x{y}^{3}-{v}^{2}{x}^{2}+{y}^{3}-{x}^{2}{y}^{2}-4 {y}^{4}-5 y{v}^{3
}+y{v}^{2}+{v}^{2}x+{y}^{3}z+2 {y}^{2}v+x{y}^{2}-5 {v}^{3}x ) 
{\alpha }^{2}\\
+ ( v+y )  ( 2 y{v}^{2}+2 {v}^{2}x+2 yvx-2 
vxz-2 zvy+{y}^{2}v-{y}^{3} ) \alpha +y ( v+y ) ^{2}
 ( x+y ) 
\end{multline}
\begin{multline}
d_z= ( 3 {v}^{2}+vx-zv-v+8 yv+yx-y-2 zy+5 {y}^{2} ) {\alpha }^{3}\\
- ( v+y )  ( 5 v-3 z+x-1+6 y ) {\alpha }^{2}+
 ( 2 {y}^{2}+2 {v}^{2}-3 zy-2 zv+4 yv ) \alpha +z ( v+y
 ) 
\end{multline}
\begin{multline}
d_v= ( x+y )  ( v+y )  ( 3 {v}^{2}+vx-zv-v+8 
yv+yx-y-2 zy+5 {y}^{2} ) {\alpha }^{3}\\
+ ( 7 zvxy-29 vx{y}^{2}
-25 {v}^{2}xy+4 yvx+4 {v}^{2}xz+4 z{v}^{2}y+3 zx{y}^{2}-4 v{x}^{
2}y\\
+6 {y}^{2}zv-{z}^{2}vx-yv{z}^{2}
-23 {y}^{3}v-22 {v}^{2}{y}^{2}-
11 x{y}^{3}-2 {v}^{2}{x}^{2}+2 {y}^{3}-2 {x}^{2}{y}^{2}\\
-8 {y}^{4}
-7 y{v}^{3}+2 y{v}^{2}+2 {v}^{2}x+2 {y}^{3}z+4 {y}^{2}v+2 x{y}^{
2}-7 {v}^{3}x ) {\alpha }^{2}\\
+ ( v+y )  ( 5 y{v}^{2}
+5 {v}^{2}x+v{x}^{2}-3 zvy+10 yvx+8 {y}^{2}v-vx-yv\\
-3 vxz-{y}^{2}-
{y}^{2}z+3 {y}^{3}+5 x{y}^{2}+{x}^{2}y-zxy-yx ) \alpha 
-v ( v+y ) ^{2} ( x+y ) 
\end{multline}
Definitions of $A_1$ and $A_2$ for eq. (\ref{r18dens}):
\begin{multline}
A_1=(2  {\alpha}-3) (32  {\alpha}^{14}+1776  {\alpha}^{13}-32304  {\alpha}^{12}\\+248136  {\alpha}^{11}-1156158  {\alpha}^{10}+3746559  
{\alpha}^9-9102790  {\alpha}^8+17374596  {\alpha}^7-26738472  {\alpha}^6\\+33372704  {\alpha}^5-33402048  {\alpha}^4+26068992  {\alpha}^3-14802944  {\alpha}^2+5259264  {\alpha}-884736)
\\+12  \sqrt{3(8  {\alpha}^9+56  {\alpha}^8-426  {\alpha}^7+940  {\alpha}^6-821  {\alpha}^5-588  {\alpha}^4+2656  {\alpha}^3-3220  {\alpha}^2+1824  {\alpha}-400)}\\
 (4  {\alpha}^4-20  {\alpha}^3+45  {\alpha}^2-16  {\alpha}-32) (4  {\alpha}^6-32  {\alpha}^5+109  {\alpha}^4-214  {\alpha}^3+284  {\alpha}^2-256  {\alpha}+128)
\end{multline}
\begin{multline}
 A_2=16  {\alpha}^{10}-576  {\alpha}^9+5048  {\alpha}^8-22816  {\alpha}^7+65969  {\alpha}^6-134476  {\alpha}^5\\+199844  {\alpha}^4-213632  {\alpha}^3+161536  {\alpha}^2-82944  {\alpha}+21504
\end{multline}


\end{document}